\documentclass[fleqn,usenatbib]{mnras}
\usepackage{newtxtext,newtxmath}
\usepackage[T1]{fontenc}
\usepackage{ae,aecompl}
\usepackage{tabularx}
\usepackage{epsfig}
\usepackage{graphicx}
\usepackage{amsmath}
\usepackage{amssymb}
\usepackage{mathrsfs}
\usepackage{hyperref}
\usepackage{mathtools}
\usepackage{afterpage}
\usepackage{epsf} 
\usepackage{caption}
\usepackage{float}
\title[Confirming new WD-ultracool dwarf binary candidates]{Confirming new white dwarf-ultracool dwarf binary candidates}

\author[Hogg et al.]{M.~A.~Hogg$^{1}$\thanks{E-mail: mah63@le.ac.uk}, S.~L. Casewell$^{1,\dagger}$, G.~A.~Wynn$^{1}$, E.~S. Longstaff$^{1,2}$, I.~P. Braker$^{1}$, 
\newauthor M.~R .Burleigh$^{1}$,  R.~H Tilbrook$^{1}$, S. Geier$^{3}$, D. Koester$^{4}$, J.~H. Debes$^{5}$, N. Lodieu$^{6,7}$ 
\\
$^{1}$Department of Physics and Astronomy, University of Leicester, Leicester, LE1 7RH, UK\\
$^{\dagger}$STFC Ernest Rutherford Fellow \\
$^{2}$Centre for Astrophysics Research, University of Hertfordshire, Hatfield AL10 9AB,UK\\
$^{3}$Institut f\"ur Physik und Astronomie, Universit\"at Potsdam, Haus 28, Karl-Liebknecht-Str. 24/25, 14476, Potsdam-Golm, Germany\\
$^{4}$Institut f\"ur Theoretische Physik und Astrophysik, Universit\"at Kiel, 24098, Kiel, Germany\\
$^{5}$Space Telescope Science Institute, Baltimore, MD 21218, USA\\
$^{6}$ Instituto de Astrof\'isica de Canarias (IAC), Calle V\'ia L\'actea s/n, E-38200 La Laguna, Tenerife, Spain \\
$^{7}$  Departamento de Astrof\'isica, Universidad de La Laguna (ULL), E-38206 La Laguna, Tenerife, Spain\\
}

\date{Received YYY; in original form ZZZ}

\pubyear{2019}

\pagerange{\pageref{firstpage}--\pageref{lastpage}}
\begin{document}
\label{firstpage}
\maketitle

\begin{abstract}
We present the results of a study to discover prospective new white dwarf-L dwarf binaries as identified by their near-infrared excesses in the UKIDSS catalogue. We obtained optical spectra to validate the white dwarf nature for 22 of the candidate primary stars, confirming ten as white dwarfs and determining their effective temperatures and gravities. For all ten white dwarfs we determined that the near-infrared excess was indeed indicative of a cool companion. Six of these are suggestive of late M dwarf companions, and three are candidate L dwarf companions, with one straddling the M$-$L boundary. We also present near-infrared spectra of eight additional candidate white dwarf-ultracool dwarf binaries where the white dwarf primary had been previously confirmed. These spectra indicate one candidate at the M$-$L boundary, three potential L dwarf companions, and one suspected M dwarf, which showed photometric variability on a $\sim$6 hour period, suggesting the system may be close. Radial velocity follow up is required to confirm whether these systems are close, or widely separated.
\end{abstract}

\begin{keywords}
stars: brown dwarfs, stars: white dwarfs, binaries: close
\end{keywords}

\section{Introduction}\label{sec:intro}
The 'brown dwarf desert' is the phenomenon where there exists a paucity of brown dwarfs orbiting within three AU of a main sequence star \citep{marcy_butler_2000, grether_lineweaver_2006, metchev_hillenbrand_2009, troup_nidever_2016}. 
A study by \cite{grether_lineweaver_2006} found two brown dwarfs from a sample of 514 stars with a companion within 10~AU. More recently \cite{triaud_amaury_2017} found a single brown dwarf within 0.2~AU from a sample of 261 FGK stars. This implies a brown dwarf companion rate of $<$1 per cent which is consistent with the previous estimates from  \citet{marcy_butler_2000} and \citet{sahlmann_segransan_2011}. To date there are 23 brown dwarfs known to transit main sequence stars with orbits within 10 AU \citep{carmichael_quinn_2020}.

The evolved form of these particular binaries  - a close white dwarf - brown dwarf (WD-BD) system - are even rarer, as the brown dwarf must survive the death of the host star \citep{green_ali_2000, farihi_becklin_2005, hoard_wachter_2007, steele_burleigh_2011, girven_gansicke_2011}. However, despite their scarcity, WD-BD binaries can offer insights into the survival of a substellar object through the common envelope phase of close binary evolution, as well as providing benchmarks for the study of irradiated atmospheres (e.g. \citealt{steele_sagalia_2013, casewell_lawrie_2015, longstaff_casewell_2017, tan_showman_2020}). 

The most commonly used method for finding these binaries is to use large catalogues of spectroscopically confirmed white dwarfs and search for a photometric near-infrared excess, indicating a brown dwarf companion. This method makes use of the fact that the white dwarf is bright in the optical wavelengths, while the much cooler brown dwarf companion emits most of its flux in the near- and mid-infrared. The use of a white dwarf to find a brown dwarf companion in this way was first attempted by \citet{probst_oconnell_1982} and \citet{probst_1983} and was successfully used to find the candidate WD-BD binary GD165 by \citet{becklin_zuckerman_1988} where the secondary was later confirmed to have a spectral type of L4 by \citet{kirkpatrick_davy_1999}. Further observations of GD165 put this object on the boundary between a very low mass star and a brown dwarf \citep{kirkpatrick_allard_1999}.

Large area optical sky surveys such as the Sloan Digital Sky Survey (SDSS: \citealt{sdss}) have catalogued hundreds of thousands of white dwarfs (e.g. \citealt{kleinman_kepler_2013, kepler_2015, kepler_2016, gentile_tremblay_2018}) which, when combined with near-infrared surveys such as the UKIRT Infrared Deep Sky Survey, (UKIDSS: \citealt{ukidds}) can be searched for brown dwarf companions (e.g. \citealt{steele_burleigh_2011, girven_gansicke_2011, verbeek_2013, rabassa_mansegas_2013}). 
The frequency of white dwarfs with brown dwarf companions is only $\approx 0.5-2$ per cent \citep{green_ali_2000, farihi_becklin_2005, hoard_wachter_2007, steele_burleigh_2011, girven_gansicke_2011}. \textit{Gaia} DR2 gives data for 486,641 candidate white dwarfs with an estimated 85 per cent completeness to 20th magnitude \citep{gaia_2018, gentile_tremblay_2018} potentially leading to the discovery of $\sim$ 2000--9500 new WD-BD systems \citep{casewell_2014}. There are however,  only 9 currently known close, detached, WD-BD binaries \citep{faraihi_christopher_2004, maxted_napiwotzki_2006, casewell_burleigh_2012, beuermann_dreizler_2013, steele_saglia_2013, farihi_parsons_2017, parsons_hermes_2017, casewell_braker_2018}.

In this paper, we present the results of a study of 30 objects identified by \citet{girven_gansicke_2011}, \citet{steele_burleigh_2011}, and \citet{verbeek_2013} as potential WD-BD binaries, all suggested to have L type secondaries. \\ 
We describe our optical spectroscopy, reduction, white dwarf fitting in Sections \ref{sec:obs}, \ref{sec:fitting}, and the near-infrared (NIR) spectroscopy and photometric excesses in Section \ref{sec:IR}. We give results on specific interesting objects in Section \ref{sec:res}, and discuss our results in \ref{sec:discussion}. All additional  figures detailing our results are in the Appendix.

\begin{table*}
\caption{Literature values, estimated via photometric fitting, for the white dwarf and brown dwarf companion for the systems with non-spectroscopically confirmed white dwarfs. The references are: 1: \citet{girven_gansicke_2011}, 2: \citet{luo_nemeth_2016}, and 3: \citet{kepler_2015} 4: \citet{verbeek_2013} 5: \citet{kepler_2016} } \label{tab:previous_knowledge} 
\begin{tabularx}{\textwidth}{l*{8}{>{\raggedleft\arraybackslash}X}}
\hline
Name & RA (hh:mm:ss) & Dec ($^{\circ}$: $'$: $''$) & Star & T$_{\rm eff}$ (K) & log \textit{g} & Companion & Ref \\ \hline
SDSS\,J010405.12$+$145906.4 & 01:04:05.12 & +14:59:07.2 & DA & 12\,000 & 7.75 & >L5 & 1 \\ 
SDSS\,J074231.98$+$285727.3 & 07:42:31.98 & +28:57:27.3 & DA & 9\,000 & 7.25 & L6 & 1 \\ 
SDSS\,J092534.99$-$014046.8 & 09:25:34.99 & $-$01:40:46.8 & DA & 14\,000 & 7.50 & >L0  & 1 \\ 
SDSS\,J095952.01$+$033032.7 & 09:59:52.01 & +03:30:32.80 & UV/Sd & 32\,000 & 8.00 & L0 & 1 \\ 
& & & SdB & 36,640 & 5.152 & L0 & 2 \\  
SDSS\,J103844.58$+$110053.5 & 10:38:44.59 & +11:00:53.55 & - & 10\,000 & 7.75 & L4 & 1 \\  
SDSS\,J133412.83$+$053415.1  & 13:34:12.83 & +05:34:15.1 & DA & 30\,000 & - & >L8 & 1 \\ 
SDSS\,J152704.26$+$080236.4 & 15:27:04.27 & +08:02:36.50 & -  & 20\,000 & 9.50 & >L8 & 1 \\
&  &  & DAZ  & 29\,560 & 5.00 & - & 3 \\  
SDSS\,J153818.86$+$064438.6 & 15:38:18.86 & +06:44:38.6 & DA & 10\,000 & 7.25 & >L5 & 1 \\ 
SDSS\,J154806.89$+$000639.4 & 15:48:06.89 & +00:06:39.4 & DA & 14\,000 & 8.50 & >L5$-$6 & 1\\ 
SDSS\,J154922.92$+$032548.6 & 15:49:22.92 & +03:25:48.6 & DA & 13\,000 & 7.25 & >L5 & 1 \\ 
SDSS\,J155128.45$-$011826.8 & 15:51:28.45 & $-$01:18:26.7 & DA & 13\,000 & 8.25 & >L8 & 1 \\ 
SDSS\,J161431.96$+$223545.0 & 16:14:31.95 & +22:35:45.0 & DA & 20\,000 & 8.5 & >L6 & 1 \\ 
%SDSSJ161942.83$+$240715.7 & 16:19:42.82 & +24:07:15.7 & DA & 9\,000 & 8.75 & >L8 & 1 \\ 
%&  &  & sdB & 27,741 & 5.227 & - & 5 \\ 
UVEX\,J184610.80$+$022032.4 & 18:46:10.80 & +02:20:32.44 & DA & 14\,000 & - & L2 & 4 \\ 
UVEX\,J185941.43$+$013954.0 & 18:59:41.43 & +01:39:53.93 & DA & 13\,000 & - & L7 & 4 \\ 
UVEX\,J191001.10$+$055542.5 & 19:10:01.08 & +05:55:42.29 & DA & 20\,000 & - & L5 & 4 \\ 
2MASS\,J20265915$+$4116436 & 20:26:59.18 & +41:16:43.87 & DA & 17\,000 & - & L3 & 4 \\ 
UVEX\,J204229.67$+$384058.0 & 20:42:29.68 & +38:40:57.95 & DA & 16\,000 & - & L2 & 4 \\ 
SDSS\,J204235.73$+$005555.8 & 20:42:35.73 & +00:55:55.8 & DA & 28\,000 & - & >L8 & 1 \\ 
UVEX\,J204856.21$+$444455.0 & 20:48:56.21 & +44:44:54.98 & DA & 16\,000 & - & L7 & 4 \\ 
SDSS\,J204922.57$-$000134.7 & 20:49:22.57 & +00:01:34.7 & DA & 16\,000 & 8.75 & >L8 & 1 \\ 
UGPS\,J210248.46$+$475058.6 & 21:02:48.44 & +47:50:58.73 & DA & 8\,000 & - & L5 & 4 \\ 
SDSS\,J211715.88$-$001548.0 & 21:17:15.88 & $-$00:15:48.0 & DA & 12\,000 & - & >L8 & 1 \\ 
%SDSSJ233040.51+145037.1 & 23:30:40.50 & +14:50:37.1 & DA & 13,000 & 9.5 & >L8 & 1 \\ 
& &  & SdB & 24\,204 & 4.968 & - & 5 \\ 
\hline
\end{tabularx}
\end{table*}

\renewcommand{\arraystretch}{1.3}
\begin{table*}
\begin{center}
\caption {The eight targets for the GNIRS NIR observations and their literature parameters The companion type is estimated from a fit to the photometric excess in the near-IR. The references are 1: \citet{eisenstein}, 2: \citet{steele_burleigh_2011}, 3: \citet{girven_gansicke_2011}, 4: \citet{kepler_2015}, 5: \citet{gaia_2018}. } \label{gnirs} 
\begin{tabular}{lcccccccc}
\hline
Name&T$_{\rm eff}$(K)&Log \textit{g} & Star & WD mass (M$_{\odot}$)& Distance (pc)&Companion Type &Ref\\
\hline 
SDSS\,J075132.52$+$200216.94&16749 $\pm$62& 8.05 $\pm$0.04& DBA& 0.54$\pm$0.02 &223$^{+8}_{-8}$&L5&3,4, 5\\
SDSS\,J090759.59$+$053638.13&19474 $\pm$293& 7.82 $\pm$0.05& DA&0.54$\pm$0.02&323 $^{+26}_{-22}$&L4,L6&1,2,3, 5\\
SDSS\,095042.31$+$011506.56& 21785 $\pm$365& 7.89 $\pm$0.06 &DA& 0.49$\pm$0.02& 321 $^{+66}_{-47}$&L8&1,3, 5\\
SDSS\,100300.08$+$093940.16&22331 $\pm$435& 7.87 $\pm$0.06 &DA& 0.57$\pm$0.03& 413 $^{+97}_{-66}$&L0&1, 2, 3, 5\\
SDSS\,J101049.48$+$040722.50&13588 $\pm$668& 7.76 $\pm$0.11 &DA& 0.48$\pm$0.05& 275 $^{+24}_{-21}$&L8&1, 3, 5\\
SDSS\,J101532.30$+$042447.66&34526 $\pm$86& 7.38 $\pm$0.07 &DA& 0.41$\pm$0.02& 792 $^{+173}_{-121}$&L4&1, 3, 5\\
SDSS\,J101607.45$+$002027.04&21045 $\pm$703& 8.48 $\pm$0.12 &DA& 0.92$\pm$0.07& 359 $^{+87}_{-59}$&L6&1, 3, 5\\
SDSS\,103736.57$+$013905.11&16300 $\pm$221& 7.73 $\pm$0.05 &DA& 0.49$\pm$0.02 &358 $^{+30}_{-26}$&M7,L5&1, 2,  3, 5\\
\hline
\end{tabular}
\end{center}
\end{table*}

\renewcommand{\arraystretch}{1}
\section{Observations and data reduction}\label{sec:obs}
\citet{steele_burleigh_2011} searched for new WD-BD binaries by looking for an infrared excess in the spectral energy distribution (SED) of spectroscopically confirmed white dwarfs, while \citet{girven_gansicke_2011} and \citet{verbeek_2013} searched for previously unidentified white dwarfs, identified by their optical broadband colours, which also appeared to have an infrared excess. \citet{girven_gansicke_2011} used SDSS photometry \citep{sdss} to find hydrogen atmosphere (DA) white dwarf candidates and then cross correlated these objects with the UKIDSS Large Area Survey \citep{ukidds} to determine if the target had a NIR excess indicative of a brown dwarf companion. The authors also identified $\approx 1$ per cent of their targets as likely having an excess that could be attributed to a dusty debris disk. Of the original data set of approximately 25\,000 objects, a subset of 153 objects were found with a candidate brown dwarf companion or disk. 

We selected all previously unobserved candidate systems from \cite{girven_gansicke_2011}, \citet{steele_burleigh_2011} and \citet{verbeek_2013} where the photometry indicated the secondary was a very low mass star or brown dwarf as indicated by an estimated spectral type of L0 or later in those papers. We obtained optical spectroscopy of systems where the white dwarf was not previously confirmed, and used a grid of pure hydrogen models to determine their effective temperatures and log \textit{g}. For systems with confirmed white dwarf primaries we obtained NIR spectroscopy to confirm the NIR excess was real and approximate the spectral type of the secondary star in the binary. Our sample is described in Table \ref{tab:previous_knowledge} for the white dwarf candidates that are photometrically identified, and Table \ref{gnirs} for the known white dwarfs.\\ 
\subsection{OSIRIS}
We used the Optical System for Imaging and low-Intermediate-Resolution Integrated Spectroscopy (OSIRIS: \citealt{cepa_2009}) instrument on the Gran Telescopio Canarias (GTC) on La Palma, to observe the 22 unconfirmed white dwarfs in the candidate WD-BD systems. The observations took place in 2015 (GTC5-15A: Lodieu) and 2017 (GTCMULTIPLE2-17A: Lodieu) using the R1000B grating, a 1.2 arcsecond slit with $2\times2$ binning which covered a wavelength range of 3630--7500$\mbox{\normalfont\AA}$ and resolution of 1018. Each object was observed with three sub-integrations of 100-300 seconds depending on target brightness which were then averaged to remove cosmic rays. We also took a spectrophotometric standard star observation for each target, and a mercury arc lamp observation for each target and standard. 

Table \ref{obs} details the dates and exposure times of the observations. 

We reduced the spectra of these objects using the standard \textsc{iraf} \citep{tody_1993} routines for long slit spectra. Each observation was bias corrected and flat-fielded before the spectrum was extracted. We used the arc lamp observation to wavelength correct the observations, and observations of the standard stars Hiltner 600, GD153, Feige 66 and Ross 640 to remove the instrument response and to flux calibrate the data. All are well used standard stars with $V$ magnitudes between 10.0 and 14.0, and typical errors of 0.05 mags. The dominant error on the flux calibration, apart from the $V$ magnitude error  is due to differences in airmass between the standard star and target which was a maximum of 0.15”.

Once the data were reduced, we inspected each spectrum to identify objects in our sample that were clearly not white dwarfs or hot subdwarfs. Of the 22 white dwarf candidates we find that UVEX\,J191001.10$+$055542.5 and UVEX\,J204856.21$+$444455.0 are emission line objects, and SDSS\,J211715.88$-$001548.0 is a hot main sequence star. Of the eight targets we observed in the NIR, the data for SDSS\,J101049.48$+$040722.50 (S/N$\sim$5) and SDSS\,J101607.45$+$002027.04 (S/N$\sim$5) were not of sufficient quality to classify the companion. 

\subsection{GNIRS}
We also observed eight previously known white dwarfs with infrared excesses indicative of an L dwarf companion (see Table \ref{gnirs}) from \citet{steele_burleigh_2011} and \citet{girven_gansicke_2011}. We selected the brightest objects with photometry suggestive of an L dwarf companion (as recorded in those papers) and obtained NIR spectra with Gemini North and the cross-dispersed spectrograph GNIRS \citep{gnirs} as part of programme GN-2017A-Q-52 (PI: Debes). We used the short camera with the 1.0" slit providing a resolution of $\approx$500 over the whole 8\,000--25\,000$\mbox{\normalfont\AA}$ spectrum. We nodded the observations with the standard 3" nod throw and 300~s exposures taken at each nod point and combined them at the reduction stage. The observations were taken in service mode over 18 nights with three hours of observations for each target, with the exception of SDSS\,J100300.08$+$093940.1622331 which was observed for four hours.

\begin{table}
\caption{Target name, date of observation, and exposure time for the 22 unconfirmed white dwarfs (above horizontal line) observed with OSIRIS in the optical and eight confirmed white dwarfs observed with GNIRS in the NIR (below line).}\label{obs} 
\begin{center}
\begin{tabular}{lccc } 
\hline
Name & Date & Airmass & Exposure\\ 
&&&time (s)\\
\hline
\noalign{\smallskip}
SDSS\,J010405.12$+$145906.4 & 2017-06-19 & 1.74 & 1150 \\
SDSS\,J074231.98$+$285727.3 & 2015-03-17 & 1.21 & 900\\
SDSS\,J092534.99$-$014046.8 & 2017-03-04 & 1.19 & 1050 \\
SDSS\,J095952.01$+$033032.7  & 2015-03-17& 1.11 & 900\\
SDSS\,J103844.58$+$110053.5 & 2015-05-19 & 1.21 & 900\\
SDSS\,J133412.83$+$053415.1 & 2017-03-04 & 1.21 & 900\\
SDSS\,J152704.26$+$080236.4 & 2015-03-18 & 1.07 & 900\\
SDSS\,J153818.86$+$064438.6 & 2017-03-04 & 1.09 & 600\\
SDSS\,J154806.89$+$000639.4 & 2017-04-01 & 1.14 & 1000\\
SDSS\,J154922.92$+$032548.6 & 2017-03-31 & 1.19 & 240\\
SDSS\,J155128.45$-$011826.8 & 2017-03-31 & 1.26 & 240\\
SDSS\,J161431.96$+$223545.0 & 2017-03-30 & 1.04 & 900 \\
%SDSS\,J161942.83+240715.7 & 2017-05-25 & 1.31 & 900\\
UVEX\,J184610.80$+$022032.4 & 2015-05-19 & 1.26 & 900\\
UVEX\,J185941.43$+$013954.0 & 2015-05-19 & 1.16 & 900\\
UVEX\,J191001.10$+$055542.5 &  2015-05-19 & 1.09 & 800\\
2MASS\,J20265915$+$4116436 &  2015-05-27 & 1.03 & 900\\
UVEX\,J204229.67$+$384058.0 &  2015-06-24 & 1.12 & 900\\
SDSS\,J204235.73$+$005555.8 & 2017-04-24 & 1.55 & 1000\\
UVEX\,J204856.21$+$444455.0 & 2015-06-15 & 1.06 &  900\\
SDSS\,J204922.57$-$000134.7 & 2017-04-23 & 1.49 & 900\\
UGPS\,J210248.46$+$475058.6 & 2015-06-23 & 1.06 & 900\\
SDSS\,J211715.88$-$001548.0 & 2017-05-03 & 1.45 & 240\\ \hline
%SDSS\,J233040.51$+$145037.1 & 2017-05-28 & 1.529 & 900\\\hline
SDSS\,J075132.52$+$200216.94 & 2017-01-11 & 1.63 & 10 x 300  \\
SDSS\,J090759.59$+$053638.13 & 2017-01-12 & 1.04 & 16 x 300\\
SDSS\,095042.31$+$011506.56 & 2017-01-15 & 1.36 & 8 x 300 \\
SDSS\,100300.08$+$093940.16 & 2017-01-16 & 1.11 & 20 x 300 \\
SDSS\,J101049.48$+$040722.50 & 2017-02-24 & 1.04 & 8 x 300 \\
SDSS\,J101532.30$+$042447.66 & 2017-03-19 & 1.05 & 8 x 300 \\
SDSS\,J101607.45$+$002027.04 & 2017-01-15 & 1.12 & 24 x 300 \\
SDSS\,103736.57$+$013905.11 & 2017-04-05 & 1.11 & 10 x 300 \\
\hline
\end{tabular}
\end{center}
\end{table}

These data were reduced using \textsc{spextool} v4.1 \citep{cushing04} which had been adapted for use with GNIRS by K. Allers (K. Allers, private comm.) and telluric corrected using \textsc{xtellcorr} \citep{vacca03}.

The spectrum of SDSS\,J101532.30$+$042447.66 has S/N$\sim$10, similar to that of most of the other objects. However, as the white dwarf in this candidate binary has a high effective temperature ($\sim$ 34\,000~K) and the suspected companion is a mid-L dwarf, the data were not of high enough S/N to determine between a lone white dwarf and a white dwarf $+$ mid-L dwarf.

\subsection{Photometry}
For two objects we also obtained optical photometry in a white light filter using the SAAO 1~m telescope with the Sutherland High-speed Optical Camera (SHOC; \citealt{coppejans}) camera. We observed SDSS\,095042.31$+$011506.5 for 2.5~hrs using 30~s exposures on the night of 2019 Feb 04 and observed SDSS\,J090759.59$+$053638.13 for 3~hrs using 180~s exposures on the night of 2019 Feb 04. The lightcurves from SHOC were reduced using the standard procedures. Bias frames were median combined and subtracted from the image frames, and twilight flats were taken, median combined and then each individual science image was divided by the combined flat. Photometry was then performed on the target and comparison stars using the starlink package \textsc{autophotom}. Apertures were fixed at twice the mean seeing (FWHM; \citealt{naylor_1998}) to minimise the impact of background noise. A clipped mean of the pixels in an annulus around the stars determined the sky background and errors was calculated from the variance in the background sky. The lightcurve of the targets was then divided by a comparison star to remove atmospheric fluctuations (e.g. from changing airmass).   
One object, SDSS\,J103736.57$+$013905.11 (EPIC251457064) was also observed by the $Kepler$ K2 mission with long cadence data. The K2 lightcurve was normalised and flagged points (such as those affected by cosmic ray hits) were removed, resulting in a lightcurve with 3566 data points.

%These objects do not feature in further analysis in this paper, leaving 19 white dwarf candidates and five previously confirmed white dwarfs with NIR excesses indicative of a cool companion.

\section{White dwarf parameters}\label{sec:fitting}
In order to measure the effective temperatures and gravities of the 19 white dwarf candidates, we created a grid of 1D model atmospheres with effective temperatures ranging from 10\,000~K to 40\,000~K in steps of 1\,000~K and log \textit{g} between 4.25 and 8.00 in steps of 0.25 dex using the 1D non-LTE model atmosphere code {\sc tlusty} \citep{hubeny1988a,hubeny1995a, TLUSTY_2017}. We then generated synthetic spectra from these model atmospheres using {\sc synspec} \citep{hubeny2011a}. The hydrogen absorption profiles for the models were calculated using the Stark broadening treatment described in \cite{tremblay2009a}.
%Objects that had low surface gravity measurements (log \textit{g} $<$6.5) are likely hot subdwarfs, and so were fitted with a more appropriate grid of models. These models are described in more detail below. 

We used the \textsc{fitsb2} software \citep{napiwotzki_yungelson_2004} to fit the effective temperature and surface gravity of each object as was done by \citet{casewell_2009}. \textsc{fitsb2} finds the best fit values of log \textit{g} and temperature through $\chi^2$ minimisation using a downhill simplex algorithm, uncertainties are found the via bootstrap method. \textsc{fitsb2} fits the individual Balmer absorption lines from H$\beta$ to H8, having retained a small amount of continuum each side of the line,  with the model grid, clipping points from the spectra that were more than 3$\sigma$ from the model to improve the fit in subsequent iterations of the fitting process. 
Our results of the fitting are in Table \ref{tab:results}. It should be noted that at effective temperatures greater than 12\,000~K, as we measure here, there is good agreement between these 1D models, and more complex 3D models such as those of \citet{tremblay_2011}. The 3D models are needed for low temperature white dwarfs to correctly models the high log \textit{g} bump caused by the mixing length creating convective overshoot. The errors given here are formal fitting errors and are likely to be an underestimate as they do not take into account the systematic uncertainties related to the instrument or reduction. \cite{tremblay_bergeron_2011} find that systematic data reduction errors and uncertainties in the physics of models give errors on the order of a few percent. We therefore use the errors as suggested by \citet*{napiwotzki_green_saffer} of 2.3 per cent in T$_{\rm eff}$ and 0.07 change in log \textit{g} in the following analysis. These results show that of the 19 white dwarf candidates, ten objects are DA white dwarfs and nine are hot subdwarfs, the expected main source of contamination for this sample. 

Once we had identified a white dwarf candidate as a hot subdwarf by a low gravity measurement, we then determined their stellar parameters using the same fitting routines as for the white dwarfs and appropriate model grids for the different classes of hot subdwarfs as described in \cite{geier_2015}. The hydrogen-rich and helium-poor (log (n(He)/n(H)) $<$ −1.0) stars with effective temperatures below 30\,000~K were fitted using a of grid of metal line blanketed LTE atmospheres with solar metallicity. The helium-poor stars with temperatures ranging from 30\,000~K to 40\,000~K were analysed using LTE models with enhanced metal line blanketing \citep{otoole06}. Metal-free NLTE models \citep{stro} were then used for hydrogen-rich stars with T$_{\rm eff}<$40\,000~K showing moderate He-enrichment (log (n(He)/n(H)) from 0.0 to $-$1.0) and for hydrogen-rich sdOs. Finally, the He-sdOs were analysed with NLTE models taking into account the line-blanketing caused by nitrogen and carbon \citep{hirsch09}. These results are in Table \ref{tab:results}.

Hot subdwarfs are thought to form from the interaction of a giant branch star with a binary companion \citep{geier_heber_2011}. However, due to their brightness and having an SED similar to that of a black body and the absence of a distance measurement, they are often found as contamination in searches for white dwarfs. These hot subdwarf systems are confirmed by their low log \textit{g} ($<$ 6.00), and large \textit{Gaia} distance ($>$1000 pc). As hot subdwarfs are thought to evolve through a common envelope phase it is not a surprise there are hot subdwarfs known with brown dwarf companions (e.g. \citealt{geier_hirsch_2011_MUCHFUSS, schaffenroth}). However, confirming a low mass companion to a hot subdwarf through an infrared excess is extremely challenging due to the low luminosity of the brown dwarf (even in the infrared) compared to that of the hot subdwarf at this distance. 
%(e.g: SDSS\,J0820$+$0008 \citep{geier_heber_2011} is located at 1513 pc \cite{gaia_2018}). 
As these objects do show an infrared excess compared to a blackbody at the subdwarf's effective temperature, this could be due to a reflection effect from a low mass companion. While we do not discuss the hot subdwarfs further within this work, lightcurves of these objects would show if they are part of a close binary system.  

\begin{table*}
\begin{center}
\caption {Our results for T$_{\rm eff}$ and log \textit{g} using \textsc{fitsb2} for the ten newly confirmed DA white dwarfs and nine hot subdwarfs. We have also included the masses and cooling times of the white dwarfs using values from the Montr\'eal white dwarf database \citep{database} and the log He/H abundances for the hot subdwarfs.} \label{tab:results} 
\begin{tabular}{llllllc}
\hline
Name & T$_{\rm eff}$(K) & Log \textit{g} & Mass (M$_{\odot}$)& Radius ($R_{\odot}$)&T$_{\rm cool}$ (Myr)& Star\\ \hline
SDSS\,J010405.12$+$145906.4 & 18\,257 $\pm$1200 & 7.52 $\pm$0.20 &0.369 $\pm$0.083&0.0176 $\pm$ $2.2\times10^{-3}$&54 $\pm$7&DA\\ 
SDSS\,J074231.98$+$285727.3 & 31\,926 $\pm$300 &7.21 $\pm$0.07 &0.403 $\pm$0.023&0.0187 $\pm$ $9\times10^{-4}$&18 $\pm$1&DA \\ 
SDSS\,J092534.99$-$014046.8 &  15\,035 $\pm$300 & 8.08 $\pm$0.07 & 0.637 $\pm$0.043 &0.0121 $\pm$ $6\times10^{-4}$ & 237 $\pm$42 & DA\\ \ 
SDSS\,J103844.58$+$110053.5 &30\,223 $\pm$500&7.45 $\pm$0.11 &0.399 $\pm$0.039&0.0186 $\pm$ $1.5\times10^{-3}$&20 $\pm$2& DA \\ 
SDSS\,J154806.89$+$000639.4 &  17\,203 $\pm$500 & 8.01 $\pm$0.07 & 0.594 $\pm$0.046&0.0127 $\pm$ $6\times10^{-4}$&127 $\pm$13 &DA \\ 
UVEX\,J184610.80$+$022032.4 & 30\,007 $\pm$300& 7.47 $\pm$0.06 &0.399 $\pm$0.020&0.0186 $\pm$ $9\times10^{-4}$&12 $\pm$1& DA \\ 
UVEX\,J185941.43$+$013954.0 & 32\,155 $\pm$300 & 7.39 $\pm$0.06 &0.404 $\pm$0.019&0.0187 $\pm$ $8\times10^{-4}$&17 $\pm$1&  DA \\ 
2MASS\,J20265915$+$4116436 &  31\,194 $\pm$200& 7.51 $\pm$0.05 &0.403 $\pm$0.016&0.0186 $\pm$ $6\times10^{-4}$&18 $\pm$1& DA \\ 
UVEX\,J204229.67$+$384058.0 &  33\,718 $\pm$200& 7.41 $\pm$0.05 &0.409 $\pm$0.021&0.0188 $\pm$ $6\times10^{-4}$&16 $\pm$1& DA \\ 
UGPS\,J210248.46$+$475058.6 &  17\,553 $\pm$400& 7.95 $\pm$0.06 &0.564 $\pm$0.037&0.0132 $\pm$ $5\times10^{-4}$&111 $\pm$7& DA \\ \hline

\hline
Name & T$_{\rm eff}$(K) & Log \textit{g} & Log He/H &&& Star\\ \hline
SDSS\,J095952.01$+$033032.7 &  34\,300 $\pm$800 & 5.45 $\pm$0.08 & $-$2.5 $\pm$ 0.2 &&&SdOB \\ 
SDSS\,J133412.83$+$053415.1  & 41\,747 $\pm$700 & 6.13 $\pm$0.17 & $-$0.56 $\pm$ 0.14&&& He-SdOB\\
SDSS\,J152704.26$+$080236.4 &  33\,700 $\pm$700 & 6.00 $\pm$0.20 & $-$2.8 $\pm$ 0.3 &&&SdB \\ 
SDSS\,J153818.86$+$064438.6 &  16\,272 $\pm$500 & 4.39 $\pm$0.08 & $<$-$3.0$ &&& BHB\\ 
SDSS\,J154922.92$+$032548.6 & 16\,805 $\pm$500 & 4.49 $\pm$0.07 & $<$-$3.0$ &&& BHB\\ 
SDSS\,J155128.45$-$011826.8 & 19\,738 $\pm$1000 & 5.01 $\pm$0.15 & $-2.34$ $\pm$0.44 &&& SdB\\ 
SDSS\,J161431.96$+$223545.0 & 28\,464 $\pm$7800 & 5.59 $\pm$0.09 & $-2.60$ $\pm$ 0.24 &&& SdB\\ 
%SDSS\,J161942.83+240715.7 & 26\,179 $\pm$672 & 5.17 $\pm$0.16 & $-2.36$ $\pm$ 0.19 &&& SdB+MS\\
SDSS\,J204235.73$+$005555.8 &  37\,412 $\pm$1900 & 5.54 $\pm$0.32 & $-0.11$ $\pm$ 0.18 &&& He-SdOB\\ 
SDSS\,J204922.57$-$000134.7 &  21\,229 $\pm$1700 & 5.52 $\pm$0.19 & $-1.87$ $\pm$ 0.16 &&& SdB\\ 
%SDSS\,J233040.51+145037.1 &  26\,387 $\pm$1429 & 5.33 $\pm$0.17 & $-2.59$ $\pm$ 0.35 &&& SdB\\ 
\hline
\end{tabular}
\end{center}
\end{table*}

\section{Searching for a NIR Excess}\label{sec:IR}
To confirm the presence of a cool companion to the DA white dwarfs now we have a measurement of effective temperature and log $g$ we followed the method used in \citet{steele_burleigh_2011} and \citet{casewell_geier_2017} of combining a \textsc{tlusty} and \textsc{synspec} DA white dwarf synthetic spectrum with ultracool dwarf template spectra of M4--L7 from the \cite[SpeX Prism Library:][]{rayner_cushing_2009}, \cite{burgasser04}, \cite{burgasser08} and \cite{Bochanski_2007}. We added the models to the template spectra by setting them both to 10~pc and using absolute magnitudes from \citet{dupuy12} for the ultracool dwarf and \citet{holberg06} for the white dwarf. It should be noted SDSS\,J075132.52$+$200216.9416749 is a DBA white dwarf, and so we used an appropriate model from \citet{koester15} which was fitted to the SDSS spectrum.

We then repeated the analysis done by \citet{girven_gansicke_2011} and \citet{verbeek_2013}. 

We refined the flux calibration of the white dwarf spectra, by fitting a polynomial to the continuum of the spectrum and dividing the spectra by this polynomial to normalise the spectrum to 1. We then multiplied by a blackbody spectrum of the appropriate effective temperature for the white dwarf.
We then took the catalogue broadband photometry from 3500--25\,000$\mbox{\normalfont\AA}$ ($ugrizJHK$) and normalised our best fitting white dwarf model and the white dwarf spectrum (from OSIRIS for the newly confirmed white dwarfs, and from SDSS for the previously identified white dwarfs) to the $r$ band. 

For the five systems where we had NIR GNIRS spectra, we stitched the optical and NIR spectra together to create a continuous spectrum. We then compared our spectra to the $ugrizJHK$ broad band photometry and the combined WD-ultracool dwarf template spectra. 
The errors on the absolute magnitudes for the NIR spectra can be $\approx$0.2--0.3 which is larger than the errors from reddening corrections which are on the order of a few percent.
The results of both the spectra and photometric fitting can be seen in Figures \ref{fig:OB14} to \ref{2102} and are summarised in Table \ref{tab:comparison} with the comparison to the literature values.

To compare our results with the scenario where there is no secondary present at all, we calculated the residuals from subtracting the model the  white dwarf alone, from the spectra. To do the same for the photometry, we created synthetic flux for each filter by convolving the white dwarf model with the relevant filter response profile. These residuals are shown on each figure as well as the residuals for the best fitting model. As can be seen in each case, the residuals in the NIR are more than 3 sigma from the white dwarf model, showing that the suggested companion is more likely to be real than a lone white dwarf.

To determine if there was a possibility that the photometric NIR excesses seen at these white dwarfs could be due to background contamination, e.g. from a red background galaxy, we searched the WISE catalogue for a detection in W1 or W2. We did not detect any of our sources. Both \citet{steele_burleigh_2011} and \citet{girven_gansicke_2011} added flags into their searches to eliminate objects that were elliptical (indicating a galaxy) or had a point spread function not indicative of a star, and so we are confident we have been able to discount any of these excesses being due to a red background galaxy.

\section{Results}\label{sec:res}
In our sample of 22 previously unconfirmed white dwarfs we spectroscopically confirm the white dwarf as the primary in ten of them. When we compare the NIR photometry for these ten white dwarfs to combined white dwarf-ultracool dwarf models we find three sources where the NIR photometry is consistent with an L dwarf companion, six which broadly match a M dwarf companion and one (UGPS\,210248.46$+$475058.6) which straddles the M-L boundary. Of the eight objects observed with GNIRS the spectrum indicates three L dwarf companions, one M dwarf companion and one (SDSS\,J100300.08$+$093940.16) which straddles the M-L boundary.

Our results show that each of these systems has either a SED which broadly matches a cool companion, or suspected photometric candidate cool companion - which is either an M or L dwarf. We are however, with one exception (SDSS\,J103736.57$+$013905.11), unable to confirm if these systems are in a close post-common envelope binary or a wider, but still unresolvable, system. There are three systems in our dataset with two epochs of $K$ band magnitudes from UKIDSS. These systems; UVEX\,J184610.80$+$022032.4, UVEX\,J185941.43$+$013954.0 and UVEX\,J204229.67$+$384058.0 all show some possible variability in the $K$ band. 

%This reflection effect is caused when a cooler star or brown dwarf is heated by a hotter, close companion, and is only seen in close binaries. All three of these systems contain M dwarf secondaries, and while the two epochs of photometry are not the same, both epochs are still broadly consistent with the spectral types given in Table \ref{tab:comparison}. These $K$ band photometry have overlapping error bars so more data are required to test whether the photometry is truly variable.

We know M dwarfs and L dwarfs can be variable due to spots, magnetic activity, clouds etc \citep{lee_berger_2010, goulding_2012, radigan_2014, buenzli_2014, vos_biller_2018} and the “variability” we see could be due to this rather than a reflection effect. A reflection effect is caused when a cooler star or brown dwarf is heated by a hotter, close companion, and is only seen in close binaries. In the case of systems with 2 $K$ band measurements, while they do differ, the errors overlap so we cannot decisively confirm they are variable. Longer monitoring would be needed to test whether the photometry is truly variable in these systems.

We present the results of specific interesting systems below.

\subsection{SDSS\,J075132.16$+$200226.8}
\begin{figure}
\begin{center}
\includegraphics[width=\columnwidth, angle=0]{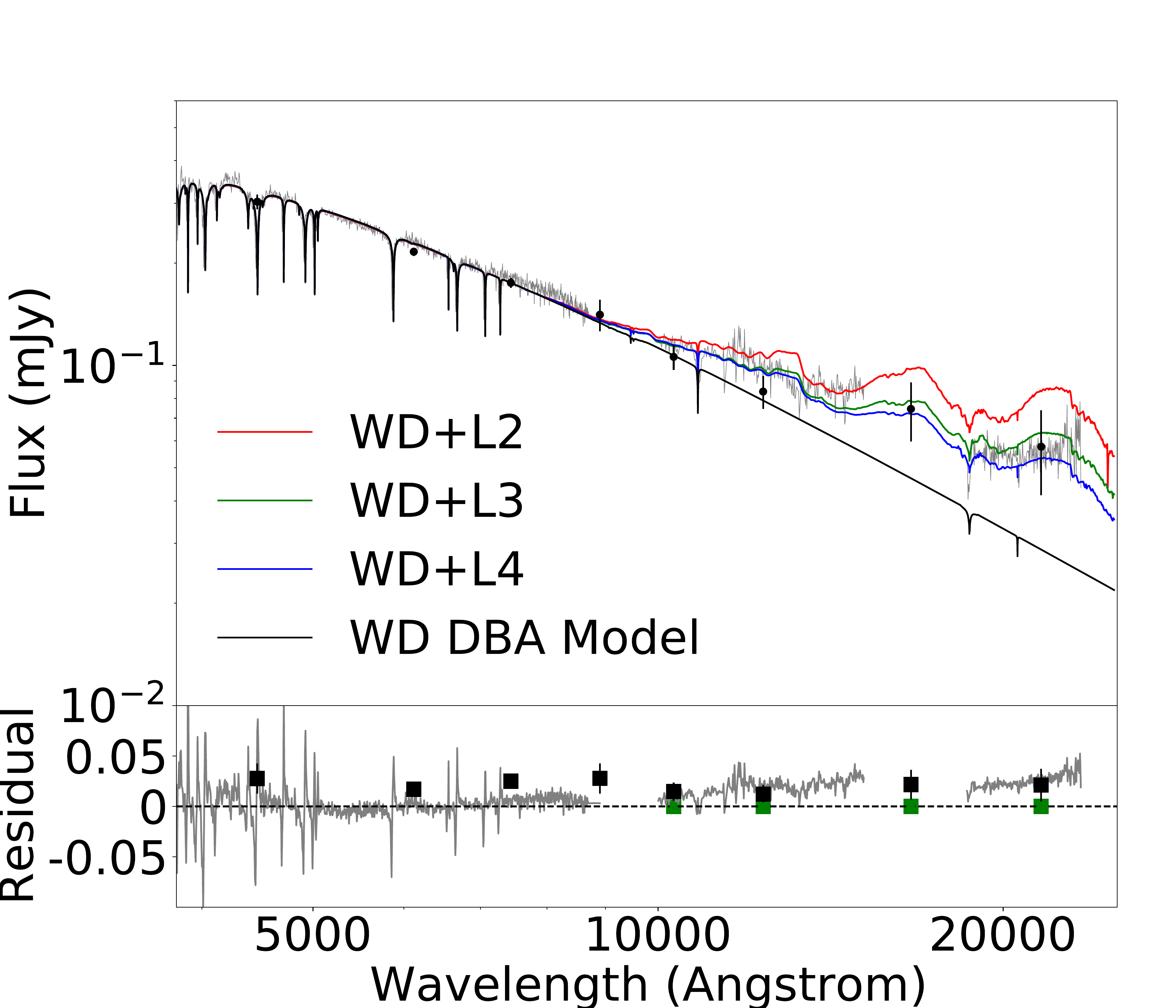}
\caption{Optical (SDSS) and GNIRS NIR spectra of SDSS\,J075132.16$+$200226.8 (S/N$\sim$13) shown with the DBA white dwarf model, and combined white dwarf+L2, white dwarf+L3 and white dwarf+L4 template spectra. The SDSS and UKIDSS photometry is also shown. Both the GNIRS spectrum and the photometric excess suggests a spectral type between L3 and L4. The residuals shown are for the fit to a WD model alone (black squares), and the combined WD-L2 dwarf template spectrum (green squares).}
\label{0751}
\end{center}
\end{figure}

\citet{kepler_2015} classified the white dwarf in this system as a DBA - although helium dominated, hydrogen is also present in the atmosphere of the white dwarf. \citet{koester15} fitted the SDSS spectrum and determined [H/He]=$-$5.17$\pm$0.25.
We used this model from \citet{koester15} to model the white dwarf, and create combined white dwarf-L dwarf models. We determine the most likely spectral type of the secondary is L3-L4 (See Fig \ref{0751}). 

If the L dwarf companion is in a close binary - it must not be close enough to cause accretion via a wind. In the weakly interacting binary NLTT5306AB the predicted upper limit to the accretion was found to be 2$\times$10$^{-15}$ M$_{\odot}$yr$^{-1}$ \citep{longstaff_casewell_2019}. This is much higher than the typical accretion rate of interstellar hydrogen assumed by \citet{koester15} of 10$^{-19}$ M$_{\odot}$yr$^{-1}$. \citet{koester15} go on to predict that DBA stars accreting at this rate will become DA white dwarfs in as little as 10$^{5}$ years. It is therefore, much more likely that this is a wide system and that the ultracool dwarf has lost very little mass, if any, to the white dwarf progenitor during the common envelope phase and that the hydrogen present in the atmosphere of the white dwarf is a residual thin layer from its previous evolution. A higher resolution spectrum than the SDSS optical spectrum shown in Fig \ref{0751} would allow more accurate abundances to be measured from the absorption features, and once a measure of the separation is obtained an accretion rate could be accurately calculated as was done for DAZ-M dwarf pairs in \citet{debes06}.

\subsection{SDSS\,J090759.59$+$053638.13}
\begin{figure}
\begin{center}
\includegraphics[width=\columnwidth, angle=0]{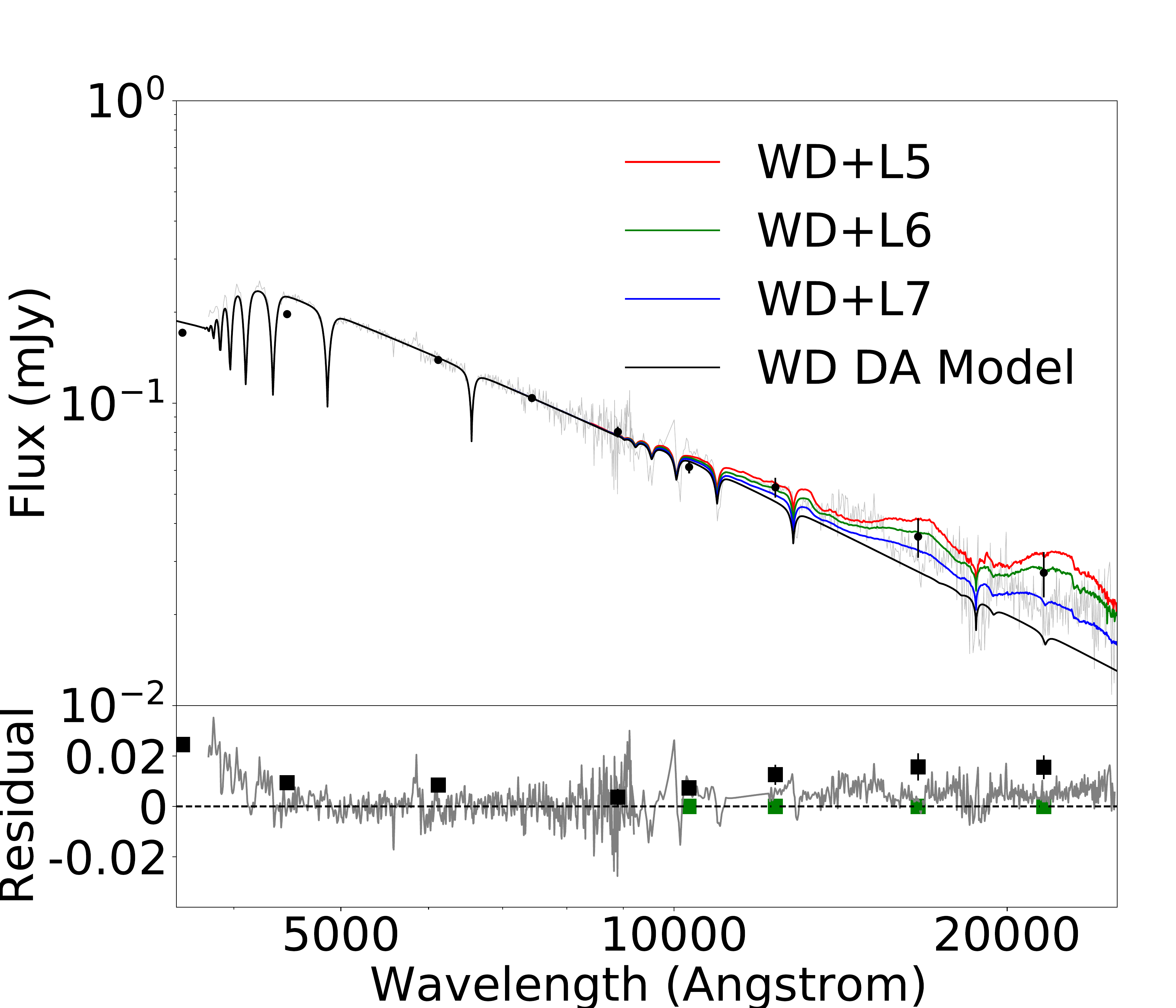}
\caption{Optical (SDSS) and GNIRS NIR spectra of SDSS\,J090759.59$+$053638.13 (S/N$\sim$9) shown with the DA white dwarf model, and combined white dwarf+L5, white dwarf+L6 and white dwarf+L7 templates. The SDSS and UKIDSS photometry is also shown. The expected companion spectral type is between L5 and L7 based on the GNIRS spectrum. The residuals shown are for the fit to a WD model alone (black squares), and the combined WD-L6 dwarf template spectrum (green squares).}
\label{0907}
\end{center}
\end{figure}

SDSS\,J090759.59$+$053638.13 is a white dwarf with a GNIRS spectrum that suggests a mid-L dwarf companion (Figure \ref{0907}). We searched for periodicity in the 3~hr SHOC lightcurve using Lomb-Scargle (LS; \citealt{Lomb1976,Scargle1982}) and discrete Fourier transform periodograms using the periodogram software packages \textsc{Vartools} \citep{hartman2016} and \textsc{Period04} \citep{lenz2005}. 
The lightcurve observed showed no variability indicative of a reflection effect or an eclipse. Binning the data to 12 minutes still did not produce any discernible reflection effect. %(Figure \ref{0907lc}), 
It may be that if this binary is close, then it has a longer period than the 3~hrs we observed, or alternatively, we may be viewing the binary at an unfavourable inclination angle that would be required to discern a large effect. More photometric data is needed before any conclusions can be drawn about this binary, although the primnary mass of 0.54$M_{\odot}$ would suggest that the system is simply wide enough for no common envelope evolution to have occurred. 

%\begin{figure}
%\begin{center}
%\includegraphics[width=0.8\columnwidth, angle=270]{0907_lc.pdf}
%\caption{SAAO lightcurve of SDSS\,J090759.59$+$053638.13 taken over three hours. No variability due to a reflection effect is detected.}
%\label{0907lc}
%\end{center}
%\end{figure}

\subsection{SDSS\,095042.31$+$011506.5}
\begin{figure}
\begin{center}
\includegraphics[width=0.8\columnwidth, angle=0]{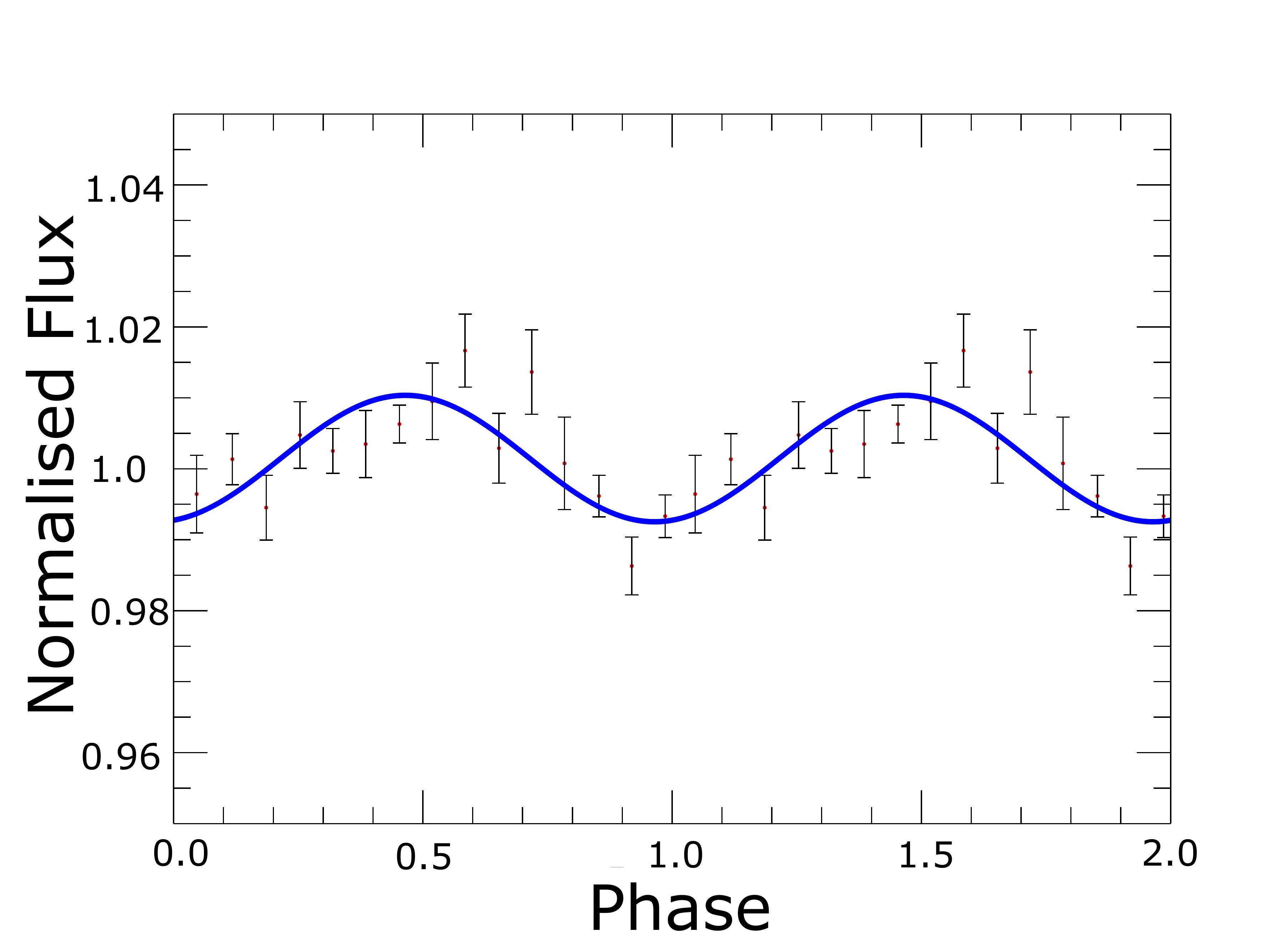}
\caption{SAAO lightcurve of SDSS\,095042.31$+$011506.5, binned to two per cent and  phase folded on the period of $\sim$ 118 min}
\label{0950lc}
\end{center}
\end{figure}

\begin{figure}
\begin{center}
\includegraphics[width=\columnwidth, angle=0]{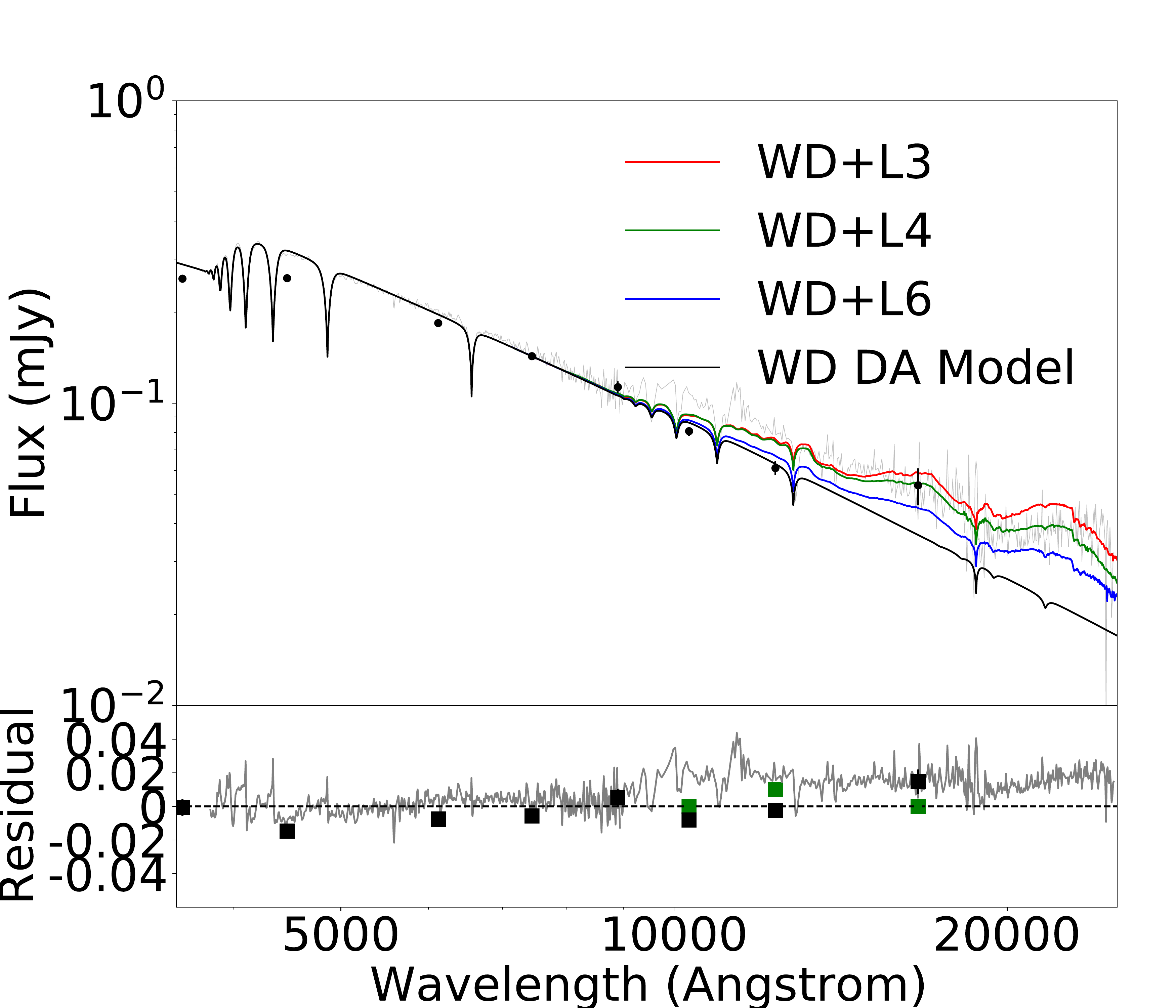}
\caption{Optical (SDSS) and GNIRS NIR spectra of SDSS\,095042.31$+$011506.56 (S/N$\sim$10) shown with the DA white dwarf model, and combined white dwarf+L3, white dwarf+L4 and white dwarf+L6 templates. The SDSS and UKIDSS photometry is also shown. The GNIRS spectrum indicates the companion broadly matches a spectral type of L4. The residuals shown are for the fit to a WD model alone (black squares), and the combined WD-L4 dwarf template spectrum (green squares).}
\label{0950}
\end{center}
\end{figure}

The white dwarf in this binary has an effective temperature of $\approx$22\,000~K, and the secondary in the system has an estimated spectral type of L4 (Figure \ref{0950}).  We searched for periodicity in the 2.5~hr SHOC lightcurve using the same method as for SDSS\,J090759.59$+$053638.13. A most likely period was identified of 0.082$\pm$0.006 days ($\approx$ 118 min) with period uncertainties determined using the bootstrap resampling (with replacement) methodology detailed in \citealt{lawrie13_1}. To determine the likelihood that the variability was due to statistical variation the peak power of Lomb Scargle (LS) spectrum was compared to a 1000 random sample lightcurves with the same mean, errors and time intervals. A false alarm probability was determined from the number of these random samples where the power of the LS spectrum exceeded that determined from the lightcurve. For the SHOC lightcurve a false alarm probability of 0.002 (0.2$\%$) was determined.  The phase folded and binned lightcurve is shown in Figure \ref{0950lc} and the variability of 0.009$\pm$0.006 per cent can clearly be seen. This amplitude of $\approx$1 per cent is similar to that seen in WD0137$-$349AB, which is consistent - SDSS\,095042.31$+$011506.5A is hotter than WD0137$-$349AB, but this binary has a slightly longer period. The data are however quite noisy, forming a broad peak in the LS periodogram indicating additional photometry is required to confirm this period.

%A most likely period was identified of 0.082$\pm$0.006 days ($\approx$ 118 min) with period uncertainties determined using the bootstrap resampling (with replacement) methodology detailed in \citealt{lawrie13_1}, and a false alarm probability of 0.002. This test identifies the peak power in the Lomb Scargle (LS) spectrum of the original data and then compares this to the peak in a 1000 random samples with the same mean, errors and time intervals.  The false alarm probability is the number of these random samples that exceed the original peak power for the actual lightcurve.

\subsection{SDSS\,J103736.57$+$013905.11}
\begin{figure}
\begin{center}
\includegraphics[width=\columnwidth, angle=0]{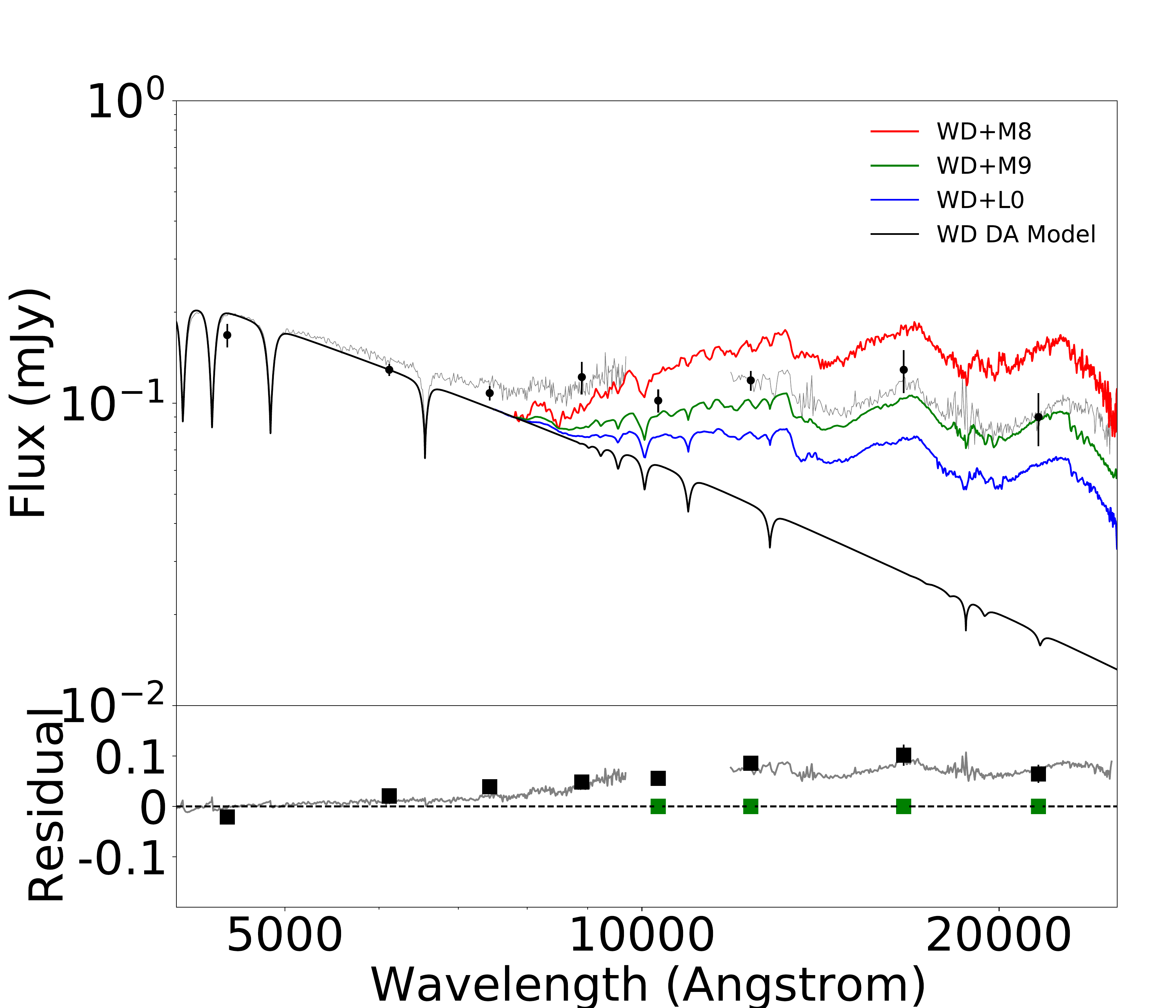}
\caption{Optical (SDSS) and GNIRS NIR spectra of SDSS\,J103736.57$+$013905.11 (S/N$\sim$23) shown with the DA white dwarf model, and combined white dwarf+M8, white dwarf+M9 and white dwarf+L0 templates. The SDSS and UKIDSS photometry is also shown. The residuals shown are for the fit to a WD model alone (black squares), and the combined WD-M9 dwarf template spectrum (green squares).}
\label{1037}
\end{center}
\end{figure}

\begin{figure}
\begin{center}
\includegraphics[width=\columnwidth, angle=0]{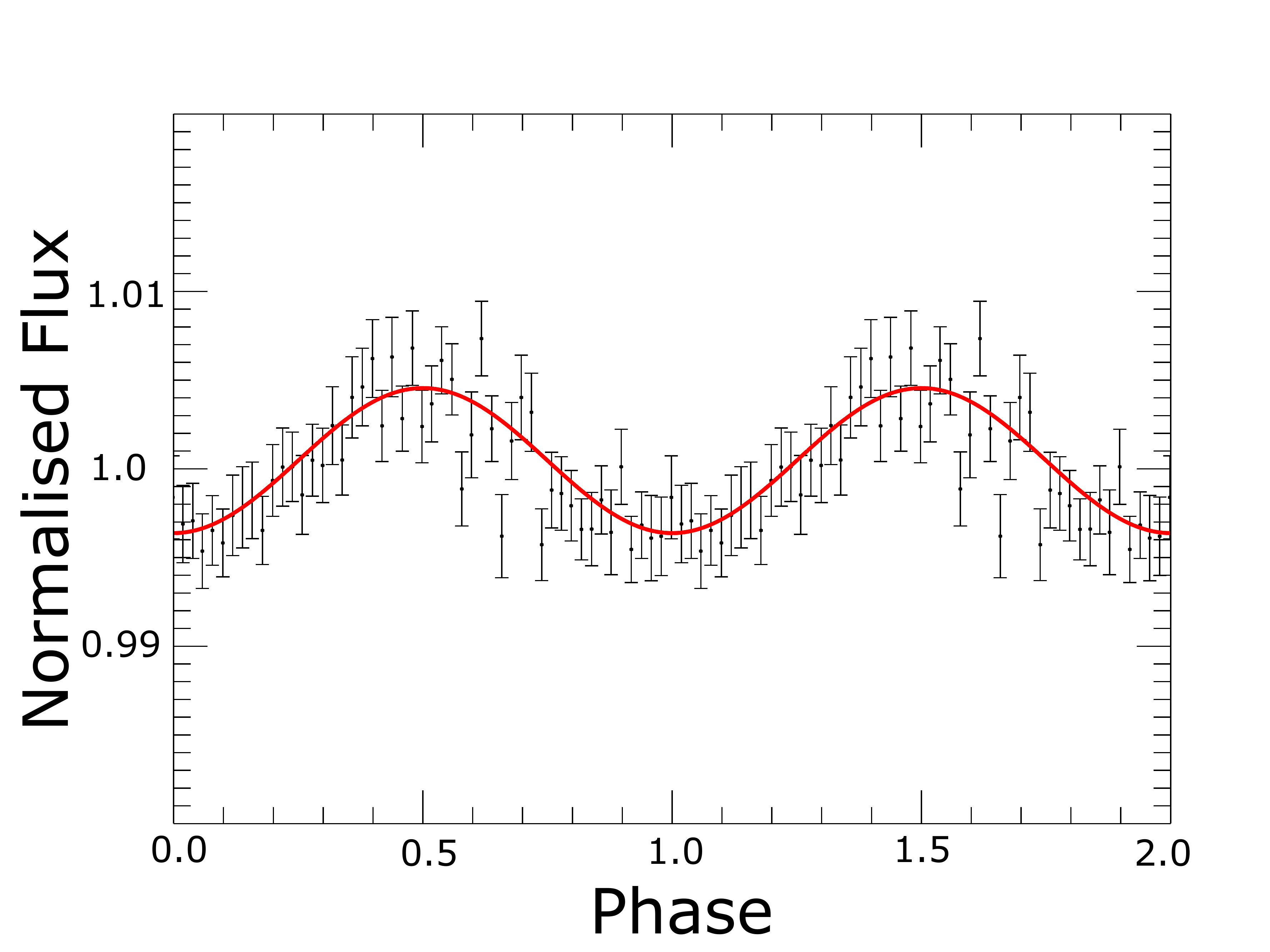}
\caption{ $Kepler$ K2 lightcurve of SDSS\,J103736.57$+$013905.11 which shows a  variability of 0.41$\pm$0.02 per cent on a period of 0.25976$\pm$0.00003 days.}
\label{k2}
\end{center}
\end{figure}

The NIR spectrum of SDSS\,J103736.57$+$013905.11 shows the secondary in this system is consistent with an M dwarf as predicted by \citet{steele_burleigh_2011} and is likely of spectral type M8-M9 (Figure \ref{1037}).  We searched for periodicity in the $Kepler$ K2 lightcurve and determined the most likely period was 0.25976$\pm$0.00003 days ($\approx$ 6.2 hrs) with period uncertainties again determined using the bootstrap resampling method. The phase folded and binned lightcurve is shown in Figure \ref{k2} and the variability of 0.41$\pm$0.02 per cent can clearly be seen.


\subsection{SDSS\,J154806.89$+$000639.4}
\begin{figure}
\begin{center}
\includegraphics[width=\columnwidth]{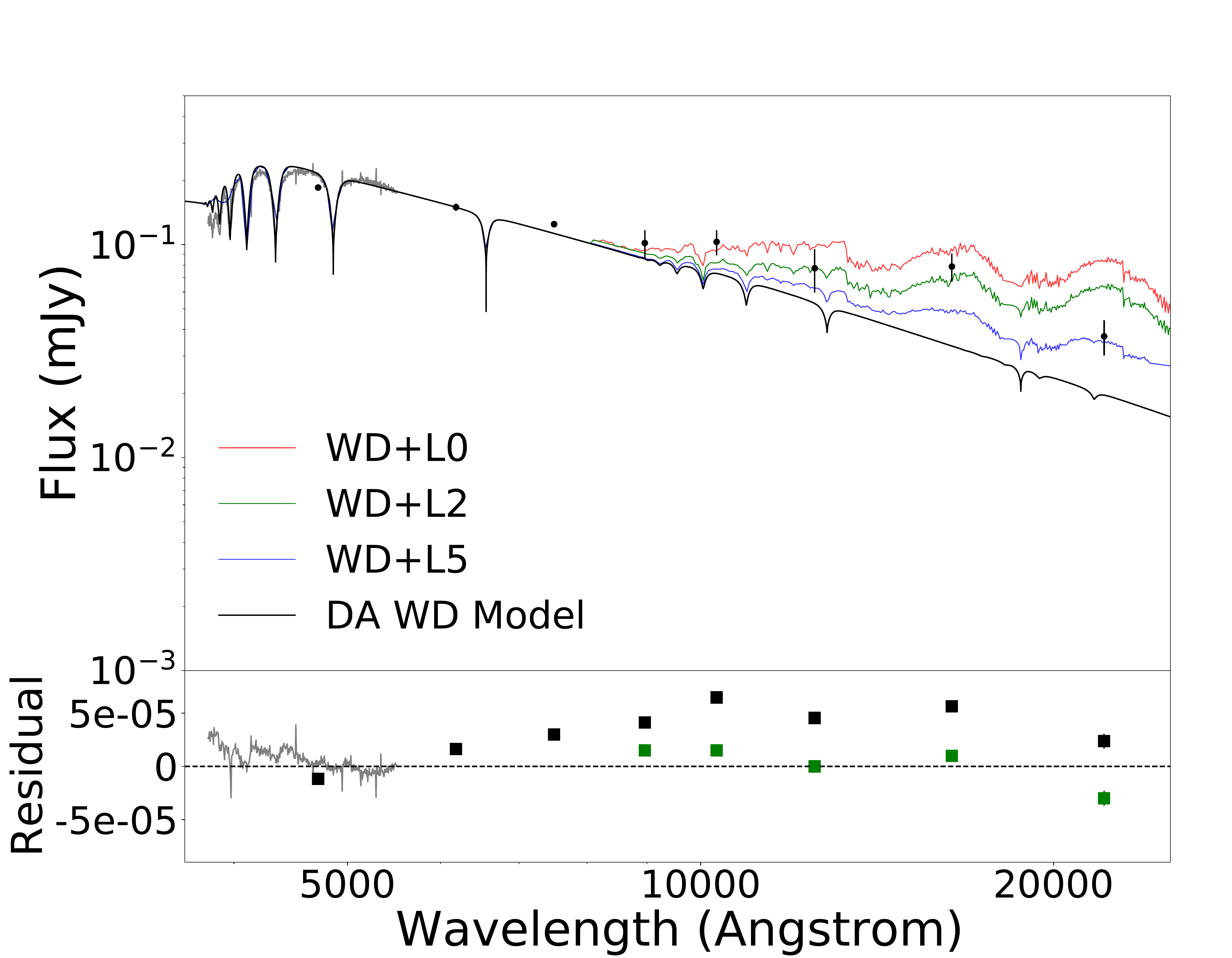}
\caption{OSIRIS spectrum of SDSS\,J154806.89$+$000639.4 shown with the SDSS and UKIDSS photometry and the DA white dwarf model combined with the template spectra of L dwarfs with spectral types between L0 and L5. The residuals shown are for the fit to a WD model alone (black squares), and the combined WD-L2 dwarf template spectrum (green squares).}
\label{fig:1548}
\end{center}
\end{figure}

SDSS\,J154806.89$+$000639.4 is an interesting system as the NIR photometry is consistent with a very wide range of spectral types (Figure \ref{fig:1548}). The $K$ band photometry suggests a spectral type significantly later than that of the $Y$, $J$ and $H$ bands. This discrepancy may indicate this is a close system with a significant reflection effect and that the $YJHK$ band data were taken at different orbital phases. Indeed, the UKIDSS $Y$ and $J$ data were taken 2005 May 08 at 13:12:48 and 13:27:25 respectively, while the $H$ and $K$ band data were taken on 2008 Mar 10 at 15:48:35 and 16:01:47 respectively. The time difference between the $H$ and $K$ data is $\approx$13 minutes and the difference between the $Y$ and $J$ data is $\approx$15 minutes.  If the spectral type of the secondary in this system is L2 as suggested by the $JH$ band data, then the $K$ band magnitude, as estimated using brown dwarf absolute magnitudes from \citet{dupuy12}, should be $\approx$ 17.45 compared to the measured value of 18.10. This would mean a decrease of 0.65 magnitudes in 15 minutes. SDSS\,J154806.89$+$000639.4 has an effective temperature of $\approx$17\,000~K, 1500~K hotter than the white dwarf in the WD0137$-$349AB system. In WD0137$-$349AB the $K$ band lightcurve changes by 0.08 mags in 15 minutes, producing a 0.3 mag difference over half the 116 min orbit. Our predicted magnitude difference for SDSS\,J154806.89$+$000639.4 is very large, but if real, suggests that this system may be close, with a short orbital period, or it may suggest that the system is eclipsing and the $K$ band data was taken during the ingress or egress of the eclipse.

\subsection{SDSS074231.98$+$285727.3}
Our fit to the optical spectrum of this object gave a temperature of ~31\,000~K, however, the photometric fit from \cite{girven_gansicke_2011} predicted a 9\,000~K temperature. The $i$ and $z$ bands are consistent with an M4 or M5 while the $YJHK$ bands are more indicative of the M5.5 or M6 template. A similar situation was highlighted by \cite{tremblay_bergeron_2011} where some objects in their survey had a mismatch between the photometric derived temperature and spectroscopically derived temperature, in some cases by as much as 70 per cent. They determined that if the white dwarf is magnetic or part of an unresolved double degenerate binary, it is possible to cause a systematic positive offset between the spectroscopic and photometric temperatures. It is also possible that a significant reflection effect may be present from the the M dwarf companion due to the high temperature white dwarf if it is part of a close binary, causing the discrepancy in spectral type. 
\begin{figure}
\includegraphics[width=\columnwidth]{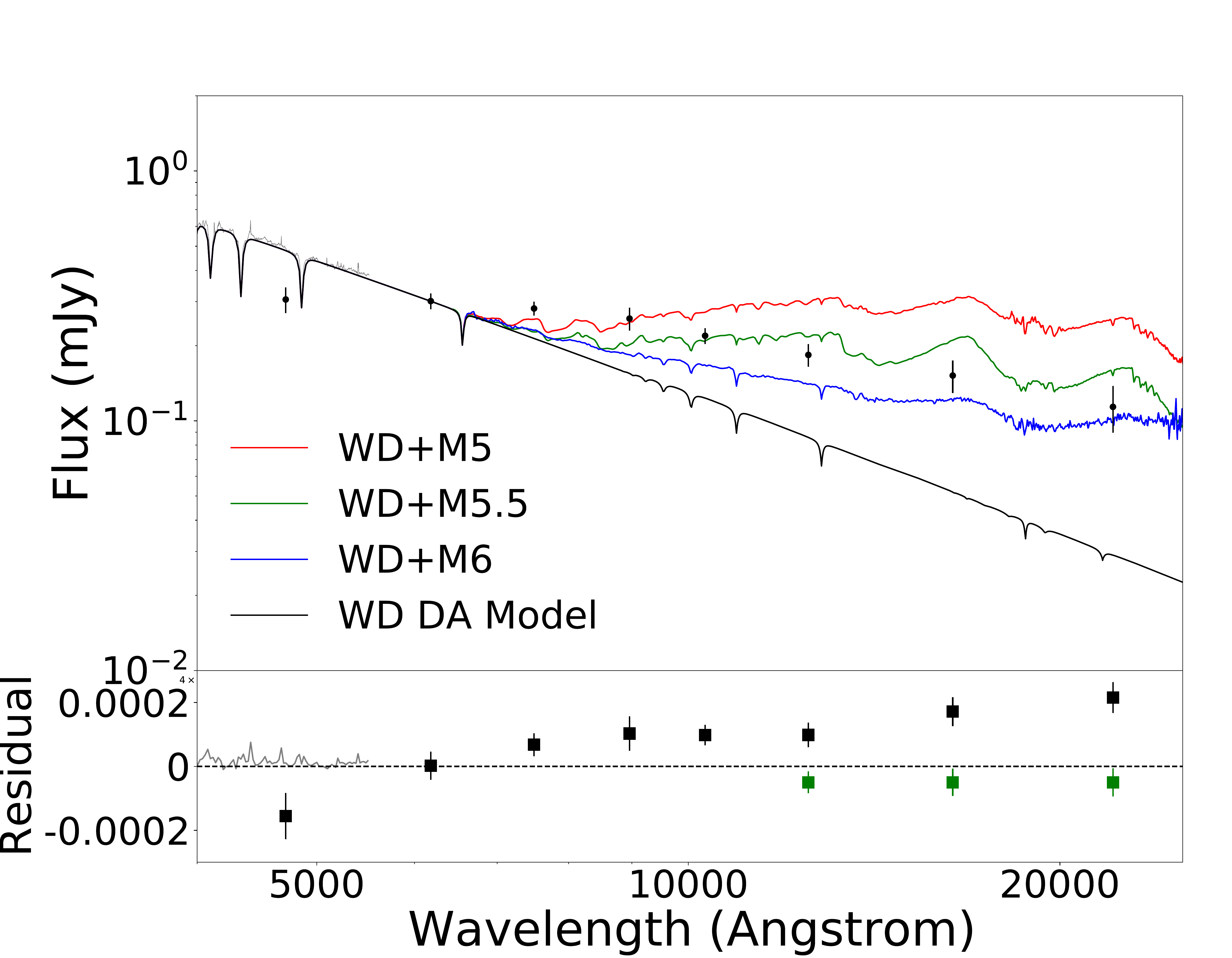}
\caption{OSIRIS spectrum of SDSS074231.98$+$285727.3 with the photometry and the combined spectra of the white dwarf model and M5, M5.5 and M6 template spectra for comparison. The data suggest that the companion is a mid-M dwarf with a spectral type between M5-M6, although the photometry do not fit one specific spectral type. The residuals shown are for the fit to a WD model alone (black squares), and the combined WD-M5.5 dwarf template spectrum (green squares).}
\label{fig:0742}
\end{figure}

\section{Discussion}\label{sec:discussion}

Our two main results are that we find that the contamination of hot subdwarfs in the sample was $\approx$50 per cent, well above the 15 per cent predicted by \cite{girven_gansicke_2011}, and that the majority of our putative secondary stars are in fact M dwarfs and not L dwarfs.

The first result - the high rate of contamination from hot subdwarfs is directly linked to the lack of distance measurements for the sample in \citet{girven_gansicke_2011}, and the lack of spectra meaning that the gravity cannot be directly measured.

The second of these two results is linked to the determination of the effective temperature by \citet{girven_gansicke_2011}.
In table \ref{tab:comparison}, we compare the effective temperature and companion spectral type to those from the literature. We find that the literature values tend to underestimate the temperature of the white dwarf by a factor of up to two. This is not hugely surprising given that \citet{girven_gansicke_2011} found a similar result when they compared their photometrically fit white dwarf effective temperatures and surface gravities to the Palomar Green sample \citep{liebert_bergeron_2005} which had an overlapping data set. They found that when using the photometry to measure the white dwarf effective temperatures they consistently underestimated the temperature compared to the results from the PG survey. They deduced that this underestimation of the temperature might be due to reddening that has not been corrected for and errors in their distance calculation. The values from \textit{Gaia} DR2 \citep{gaia_2018} which were not available to \citet{girven_gansicke_2011} provide accurate distances, and allow hot subdwarfs to be removed from the sample without the need for spectra.
Studies by \cite{tremblay_bergeron_2011} and \cite{ genest-beaulieu_bergeron_2015} compared parameters obtained from either photometric or spectroscopic fitting and found that the two broadly match in log \textit{g}. The authors also found that the temperatures  from the the photometric and spectroscopic fits matched well, however they found a small but significant temperature offset above $T_{eff} ~14\,000~K$ where spectroscopic determinations were systematically higher than those from a photometric fit by $\approx$2\%.

\begin{table*}
\begin{center}
\caption {Comparison of literature temperatures and companion spectral types to our results. The temperature comparison is only performed for the objects we obtained optical spectra for. The Phot/Spec column details whether our companion spectral type is based on NIR spectra or photometry. The references are:  1: \citet{girven_gansicke_2011}, 2: \citet{steele_burleigh_2011}, 3: \citet{eisenstein}, 4: \citet{kepler_2015}, 5: \citet{verbeek_2013}. } \label{tab:comparison} 
\begin{tabular}{lcccccc }
\hline
Name & T$_{\rm eff}$(K) & T$_{\rm eff}$ (K) & Companion & Companion & Phot/Spec & Ref\\ 
&FITSB2&Literature&&Literature &\\\hline
SDSS\,J010405.12$+$145906.4 & 18\,257 $\pm$1200 & 12\,000 & L0$-$L2 & L5 &Phot & 1\\ 
SDSS\,J074231.98$+$285727.3 & 31\,926 $\pm$300 & 9\,000 & M5$-$M6 & L6& Phot & 1\\ 
SDSS\,J075132.16$+$200226.8 & 16\,749 $\pm$100& 20\,000 & L3$-$L4 & L5 &Spec & 1,4\\ 
SDSS\,J090759.59$+$053638.13&-&-&L5-L7&L4,L6 &Spec & 1, 2, 3, \\
SDSS\,J092534.99$-$014046.8  & 15\,035 $\pm$400 & 14\,000 & L1$-$L2 & >L0 & Phot & 1\\ 
SDSS\, 095042.31$+$011506.56&-&-&L4&L8 & Spec & 1,3\\
SDSS\, 100300.08$+$093940.16&-&-&M8$-$L0&L0 & Spec & 1, 2, 3,\\
SDSS\, 103736.57$+$013905.11&-&-&M8$-$M9&M7,L5 & Spec & 1, 2, 3\\
SDSS\,J103844.58$+$110053.5 & 30\,223 $\pm$500& 10\,000 & M5$-$M6 & L4 & Phot & 1\\ 
SDSS\,J154806.89$+$000639.4 & 17\,203 $\pm$500 & 14\,000 & L0$-$L5 & >L5$-$L6 & Phot & 1\\\ 
UVEX\,J184610.80$+$022032.4 & 30\,007 $\pm$300 & 14\,000 & M5$-$M6 & L2 & Phot & 5 \\ 
UVEX\,J185941.43$+$013954.0 & 32\,155 $\pm$300 & 13\,000 & M5$-$M6 & L7 & Phot & 5\\ 
2MASS\,J20265915$+$4116436 & 31\,194 $\pm$200& 17\,000 & M5 & L3 & Phot & 5\\
UVEX\,J204229.67$+$384058.0 & 33\,718 $\pm$200& 16\,000 & M5$-$M6 & L2 & Phot & 5\\ 
UGPS\,J210248.46$+$475058.6 & 17\,553 $\pm$400& 8\,000 & M9$-$L1 & L5 & Phot & 5\\ \hline
\end{tabular}
\end{center}
\end{table*}

Despite determining the putative secondary in these systems is more likely to be an M dwarf than was previously predicted for our sample, all of the the newly-confirmed white dwarfs appear to have photometry consistent with M dwarf or L dwarf companions and so we find the method of \citet{girven_gansicke_2011} and \citet{steele_burleigh_2011} is successful in identifying candidate white dwarf-ultracool dwarf binaries. Caution should be used when using solely photometric methods   
as the effective temperatures are likely to be underestimated, leading to  predictions of later spectral type companions than are actually present. If parallaxes (e.g. from $Gaia$, \citealt{gaia_2018}) are not used then there is also likely to be a high contamination rate from hot subdwarfs. Using distances from \textit{Gaia} combined with SED fitting will allow more accurate effective temperatures to be derived, resulting in better estimates of the spectral type of the secondary (e.g. the $Gaia$ catalogue of \citealt{gentile_tremblay_2018}). 

We have identified six candidate white dwarf-L dwarf binaries (three with NIR spectra). Specifically we found one candidate binary, SDSS\,J090759.59$+$053638.13, that has an approximate spectral companion type of L5-L7. The lightcurve of this binary showed that it is unlikely to have an eclipsing orbit $<$3hrs but additional data will reveal if it is a close binary and a higher quality spectrum will reveal more about the nature of the secondary.  

We note that our systems have a total age of $\approx$1--10~Gyrs. At this age, a 70 M$_{\rm Jup}$ object has spectral type later than L2, using the DUSTY models of  \citet{baraffe_chabrier_2001}, therefore it is likely that some of these L dwarf companions are genuine brown dwarfs.

Lightcurves of these WD-L dwarf binary systems should reveal (via a reflection effect) if they are close binaries, and therefore suitable for radial velocity follow up. One potentially interesting result is that of our newly confirmed DA white dwarfs, seven have masses of $\sim$0.4~M$_{\odot}$, lower than the mean white dwarf mass of 0.6~M$_{\odot}$ \citep{tremblay19}. It is thought that in order to form a white dwarf with such a low mass, a phase of close binary interaction while on the red giant branch must have occurred \citep{marsh95}, strengthening the case that these binaries may be close, post-common envelope systems. These new binaries, if close, will increase the number of known close, detached WD-L dwarf systems, providing a wider variety of white dwarf effective temperatures and L dwarf spectral types to study irradiated L dwarf atmospheres. A wider range of secondary masses in these binaries will also allow tighter limits to be put on the masses of objects that can survive the common envelope phase of evolution.

\section{Conclusions}
We have confirmed ten new white dwarfs, all of which have likely M or L dwarf companions. We have also confirmed nine new hot subdwarfs - the main source of contamination in the sample. We also obtained NIR spectra for eight confirmed white dwarfs with suspected companions and find five which are consistent with having a cool companion. Photometry suggests one system may be a close binary in a $\sim$ 6 hour period, and another we have tentatively identified to have a $\sim$ 118 min period.

\section{Acknowledgements}
We thank our referee, Stuart Littlefair, for the valuable comments on the manuscript.
We would like to thank Katelyn Allers for kindly providing the GNIRS reduction tool to be used within Spextool and Ry Cutter for their helpful discussion and proofreading of the manuscript. SLC acknowledges support from an STFC Ernest Rutherford Fellowship. MAH, RHT and ESL acknowledges support from STFC studentships and IPB acknowledges support from the College of Science and Engineering at the University of Leicester.

This work is based on observations made with the Gran Telescopio Canarias (GTC), installed in the Spanish Observatorio del Roque de los Muchachos of the Instituto de Astrofisica de Canarias, in the island of La Palma. Also based on observations obtained at the Gemini Observatory, which is operated by the Association of Universities for Research in Astronomy, Inc., under a cooperative agreement with the NSF on behalf of the Gemini partnership: the National Science Foundation (United States), National Research Council (Canada), CONICYT (Chile), Ministerio de Ciencia, Tecnolog\'{i}a e Innovaci\'{o}n Productiva (Argentina), Minist\'{e}rio da Ci\^{e}ncia, Tecnologia e Inova\c{c}\~{a}o (Brazil), and Korea Astronomy and Space Science Institute (Republic of Korea). This work has also made use of the Montreal White Dwarf Database which is produced and maintained by Prof. Patrick Dufour at Universit\'e de Montr\'eal.

\section{Data availability}
The data underlying this article were accessed from   \href{http://gtc.sdc.cab.inta-csic.es/gtc/index.jsp}{Gran Telescopio Canarias} (GTC5-15A: Lodieu and GTC-17A: Lodieu), \href{https://archive.gemini.edu/searchform/}{Gemini North} (GN-2017A-Q-52: Debes) and The South African Astronomical Observatory (PI: Burleigh). The derived data generated in this research will be shared on request to the corresponding author. The data reduction pipelines and white dwarf models are all open source. The Ultracool dwarf template spectra are available from the Spex Prism Library.

\bibliographystyle{mnras}
\bibliography{MahPapers} 

\begin{thebibliography}{}
\makeatletter
\relax
\def\mn@urlcharsother{\let\do\@makeother \do\$\do\&\do\#\do\^\do\_\do\%\do\~}
\def\mn@doi{\begingroup\mn@urlcharsother \@ifnextchar [ {\mn@doi@}
  {\mn@doi@[]}}
\def\mn@doi@[#1]#2{\def\@tempa{#1}\ifx\@tempa\@empty \href
  {http://dx.doi.org/#2} {doi:#2}\else \href {http://dx.doi.org/#2} {#1}\fi
  \endgroup}
\def\mn@eprint#1#2{\mn@eprint@#1:#2::\@nil}
\def\mn@eprint@arXiv#1{\href {http://arxiv.org/abs/#1} {{\tt arXiv:#1}}}
\def\mn@eprint@dblp#1{\href {http://dblp.uni-trier.de/rec/bibtex/#1.xml}
  {dblp:#1}}
\def\mn@eprint@#1:#2:#3:#4\@nil{\def\@tempa {#1}\def\@tempb {#2}\def\@tempc
  {#3}\ifx \@tempc \@empty \let \@tempc \@tempb \let \@tempb \@tempa \fi \ifx
  \@tempb \@empty \def\@tempb {arXiv}\fi \@ifundefined
  {mn@eprint@\@tempb}{\@tempb:\@tempc}{\expandafter \expandafter \csname
  mn@eprint@\@tempb\endcsname \expandafter{\@tempc}}}

\bibitem[\protect\citeauthoryear{{Abazajian} et~al.,}{{Abazajian}
  et~al.}{2009}]{sdss}
{Abazajian} K.~N.,  et~al., 2009, \mn@doi [\apjs]
  {10.1088/0067-0049/182/2/543}, \href
  {http://adsabs.harvard.edu/abs/2009ApJS..182..543A} {182, 543}

\bibitem[\protect\citeauthoryear{{Baraffe}, {Chabrier}, {Allard}  \&
  {Hauschildt}}{{Baraffe} et~al.}{2002}]{baraffe_chabrier_2001}
{Baraffe} I.,  {Chabrier} G.,  {Allard} F.,   {Hauschildt} P.~H.,  2002,
  \mn@doi [\aap] {10.1051/0004-6361:20011638}, \href
  {https://ui.adsabs.harvard.edu/abs/2002A&A...382..563B} {382, 563}

\bibitem[\protect\citeauthoryear{{Becklin} \& {Zuckerman}}{{Becklin} \&
  {Zuckerman}}{1988}]{becklin_zuckerman_1988}
{Becklin} E.~E.,  {Zuckerman} B.,  1988, \mn@doi [\nat] {10.1038/336656a0},
  \href {https://ui.adsabs.harvard.edu/abs/1988Natur.336..656B} {336, 656}

\bibitem[\protect\citeauthoryear{{Beuermann} et~al.,}{{Beuermann}
  et~al.}{2013}]{beuermann_dreizler_2013}
{Beuermann} et~al., 2013, \mn@doi [A\&A] {10.1051/0004-6361/201322241}, 558,
  A96

\bibitem[\protect\citeauthoryear{Bochanski, West, Hawley  \& Covey}{Bochanski
  et~al.}{2007}]{Bochanski_2007}
Bochanski J.~J.,  West A.~A.,  Hawley S.~L.,   Covey K.~R.,  2007, \mn@doi [The
  Astronomical Journal] {10.1086/510240}, 133, 531

\bibitem[\protect\citeauthoryear{Buenzli, Apai, Radigan, Reid  \&
  Flateau}{Buenzli et~al.}{2014}]{buenzli_2014}
Buenzli E.,  Apai D.,  Radigan J.,  Reid I.~N.,   Flateau D.,  2014, \mn@doi
  [The Astrophysical Journal] {10.1088/0004-637x/782/2/77}, 782, 77

\bibitem[\protect\citeauthoryear{{Burgasser}, {McElwain}, {Kirkpatrick},
  {Cruz}, {Tinney}  \& {Reid}}{{Burgasser} et~al.}{2004}]{burgasser04}
{Burgasser} A.~J.,  {McElwain} M.~W.,  {Kirkpatrick} J.~D.,  {Cruz} K.~L.,
  {Tinney} C.~G.,   {Reid} I.~N.,  2004, \mn@doi [\aj] {10.1086/383549}, \href
  {http://adsabs.harvard.edu/abs/2004AJ....127.2856B} {127, 2856}

\bibitem[\protect\citeauthoryear{{Burgasser}, {Liu}, {Ireland}, {Cruz}  \&
  {Dupuy}}{{Burgasser} et~al.}{2008}]{burgasser08}
{Burgasser} A.~J.,  {Liu} M.~C.,  {Ireland} M.~J.,  {Cruz} K.~L.,   {Dupuy}
  T.~J.,  2008, \mn@doi [\apj] {10.1086/588379}, \href
  {http://adsabs.harvard.edu/abs/2008ApJ...681..579B} {681, 579}

\bibitem[\protect\citeauthoryear{{Carmichael} et~al.,}{{Carmichael}
  et~al.}{2020}]{carmichael_quinn_2020}
{Carmichael} T.~W.,  et~al., 2020, arXiv e-prints, \href
  {https://ui.adsabs.harvard.edu/abs/2020arXiv200201943C} {p. arXiv:2002.01943}

\bibitem[\protect\citeauthoryear{{Casewell}}{{Casewell}}{2014}]{casewell_2014}
{Casewell} S.~L.,  2014, \memsai, \href
  {http://adsabs.harvard.edu/abs/2014MmSAI..85..731C} {85, 731}

\bibitem[\protect\citeauthoryear{{Casewell}, {Dobbie}, {Napiwotzki},
  {Burleigh}, {Barstow}  \& {Jameson}}{{Casewell} et~al.}{2009}]{casewell_2009}
{Casewell} S.~L.,  {Dobbie} P.~D.,  {Napiwotzki} R.,  {Burleigh} M.~R.,
  {Barstow} M.~A.,   {Jameson} R.~F.,  2009, \mn@doi [\mnras]
  {10.1111/j.1365-2966.2009.14593.x}, \href
  {http://adsabs.harvard.edu/abs/2009MNRAS.395.1795C} {395, 1795}

\bibitem[\protect\citeauthoryear{Casewell et~al.,}{Casewell
  et~al.}{2012}]{casewell_burleigh_2012}
Casewell S.~L.,  et~al., 2012, The Astrophysical Journal Letters, 759, L34

\bibitem[\protect\citeauthoryear{Casewell et~al.,}{Casewell
  et~al.}{2015}]{casewell_lawrie_2015}
Casewell S.~L.,  et~al., 2015, \mn@doi [Monthly Notices of the Royal
  Astronomical Society] {10.1093/mnras/stu2721}, 447, 3218

\bibitem[\protect\citeauthoryear{{Casewell}, {Geier}  \& {Lodieu}}{{Casewell}
  et~al.}{2017}]{casewell_geier_2017}
{Casewell} S.~L.,  {Geier} S.,   {Lodieu} N.,  2017, in {Tremblay} P.-E.,
  {Gaensicke} B.,   {Marsh} T.,  eds,  Astronomical Society of the Pacific
  Conference Series Vol. 509, 20th European White Dwarf Workshop. p.~133

\bibitem[\protect\citeauthoryear{{Casewell} et~al.,}{{Casewell}
  et~al.}{2018}]{casewell_braker_2018}
{Casewell} S.~L.,  et~al., 2018, \mn@doi [\mnras] {10.1093/mnras/sty245}, \href
  {http://adsabs.harvard.edu/abs/2018MNRAS.476.1405C} {476, 1405}

\bibitem[\protect\citeauthoryear{{Cepa}}{{Cepa}}{2009}]{cepa_2009}
{Cepa} J.,  2009, in IAC Talks, Astronomy and Astrophysics Seminars from the
  Instituto de Astrof{\'{\i}}sica de Canarias. p.~142

\bibitem[\protect\citeauthoryear{{Coppejans} et~al.,}{{Coppejans}
  et~al.}{2013}]{coppejans}
{Coppejans} R.,  et~al., 2013, \mn@doi [\pasp] {10.1086/672156}, \href
  {http://adsabs.harvard.edu/abs/2013PASP..125..976C} {125, 976}

\bibitem[\protect\citeauthoryear{{Cushing}, {Vacca}  \& {Rayner}}{{Cushing}
  et~al.}{2004}]{cushing04}
{Cushing} M.~C.,  {Vacca} W.~D.,   {Rayner} J.~T.,  2004, \mn@doi [\pasp]
  {10.1086/382907}, \href {http://adsabs.harvard.edu/abs/2004PASP..116..362C}
  {116, 362}

\bibitem[\protect\citeauthoryear{{Debes}}{{Debes}}{2006}]{debes06}
{Debes} J.~H.,  2006, \mn@doi [\apj] {10.1086/508132}, \href
  {https://ui.adsabs.harvard.edu/abs/2006ApJ...652..636D} {652, 636}

\bibitem[\protect\citeauthoryear{{Dufour}, {Blouin}, {Coutu},
  {Fortin-Archambault}, {Thibeault}, {Bergeron}  \& {Fontaine}}{{Dufour}
  et~al.}{2017}]{database}
{Dufour} P.,  {Blouin} S.,  {Coutu} S.,  {Fortin-Archambault} M.,  {Thibeault}
  C.,  {Bergeron} P.,   {Fontaine} G.,  2017, in {Tremblay} P.-E.,  {Gaensicke}
  B.,   {Marsh} T.,  eds,  Astronomical Society of the Pacific Conference
  Series Vol. 509, 20th European White Dwarf Workshop. p.~3 (\mn@eprint {arXiv}
  {1610.00986})

\bibitem[\protect\citeauthoryear{{Dupuy} \& {Liu}}{{Dupuy} \&
  {Liu}}{2012}]{dupuy12}
{Dupuy} T.~J.,  {Liu} M.~C.,  2012, \mn@doi [\apjs]
  {10.1088/0067-0049/201/2/19}, \href
  {http://adsabs.harvard.edu/abs/2012ApJS..201...19D} {201, 19}

\bibitem[\protect\citeauthoryear{{Eisenstein} et~al.,}{{Eisenstein}
  et~al.}{2006}]{eisenstein}
{Eisenstein} D.~J.,  et~al., 2006, \mn@doi [\apjs] {10.1086/507110}, \href
  {http://adsabs.harvard.edu/abs/2006ApJS..167...40E} {167, 40}

\bibitem[\protect\citeauthoryear{{Elias}, {Joyce}, {Liang}, {Muller}, {Hileman}
   \& {George}}{{Elias} et~al.}{2006}]{gnirs}
{Elias} J.~H.,  {Joyce} R.~R.,  {Liang} M.,  {Muller} G.~P.,  {Hileman} E.~A.,
   {George} J.~R.,  2006, in Society of Photo-Optical Instrumentation Engineers
  (SPIE) Conference Series. p. 62694C, \mn@doi{10.1117/12.671817}

\bibitem[\protect\citeauthoryear{Farihi \& Christopher}{Farihi \&
  Christopher}{2004}]{faraihi_christopher_2004}
Farihi J.,  Christopher M.,  2004, The Astronomical Journal, 128, 1868

\bibitem[\protect\citeauthoryear{{Farihi}, {Becklin}  \& {Zuckerman}}{{Farihi}
  et~al.}{2005}]{farihi_becklin_2005}
{Farihi} J.,  {Becklin} E.~E.,   {Zuckerman} B.,  2005, \mn@doi [\apjs]
  {10.1086/444362}, \href
  {https://ui.adsabs.harvard.edu/abs/2005ApJS..161..394F} {161, 394}

\bibitem[\protect\citeauthoryear{{Farihi}, {Parsons}  \&
  {G{\"a}nsicke}}{{Farihi} et~al.}{2017}]{farihi_parsons_2017}
{Farihi} J.,  {Parsons} S.~G.,   {G{\"a}nsicke} B.~T.,  2017, \mn@doi [Nature
  Astronomy] {10.1038/s41550-016-0032}, \href
  {http://adsabs.harvard.edu/abs/2017NatAs...1E..32F} {1, 0032}

\bibitem[\protect\citeauthoryear{{Gaia Collaboration}}{{Gaia
  Collaboration}}{2018}]{gaia_2018}
{Gaia Collaboration} 2018, VizieR Online Data Catalog, \href
  {http://adsabs.harvard.edu/abs/2018yCat.1345....0G} {1345}

\bibitem[\protect\citeauthoryear{{Geier} et~al.,}{{Geier}
  et~al.}{2011a}]{geier_hirsch_2011_MUCHFUSS}
{Geier} S.,  et~al., 2011a, \mn@doi [\aap] {10.1051/0004-6361/201015316}, \href
  {http://adsabs.harvard.edu/abs/2011A%26A...530A..28G} {530, A28}

\bibitem[\protect\citeauthoryear{{Geier} et~al.,}{{Geier}
  et~al.}{2011b}]{geier_heber_2011}
{Geier} S.,  et~al., 2011b, in {Schuh} S.,  {Drechsel} H.,   {Heber} U.,  eds,
  American Institute of Physics Conference Series Vol. 1331, American Institute
  of Physics Conference Series. pp 163--169 (\mn@eprint {arXiv} {1012.3839}),
  \mn@doi{10.1063/1.3556196}

\bibitem[\protect\citeauthoryear{{Geier} et~al.,}{{Geier}
  et~al.}{2015}]{geier_2015}
{Geier} S.,  et~al., 2015, \mn@doi [\aap] {10.1051/0004-6361/201525666}, \href
  {http://adsabs.harvard.edu/abs/2015A%26A...577A..26G} {577, A26}

\bibitem[\protect\citeauthoryear{{Genest-Beaulieu}, {Bergeron}  \&
  {Darveau-Bernier}}{{Genest-Beaulieu}
  et~al.}{2015}]{genest-beaulieu_bergeron_2015}
{Genest-Beaulieu} C.,  {Bergeron} P.,   {Darveau-Bernier} A.,  2015,
  {Comparative Analysis of Atmospheric Parameters Obtained from the Photometric
  and Spectroscopic Techniques}.
p.~45

\bibitem[\protect\citeauthoryear{{Gentile Fusillo} et~al.,}{{Gentile Fusillo}
  et~al.}{2019}]{gentile_tremblay_2018}
{Gentile Fusillo} N.~P.,  et~al., 2019, \mn@doi [\mnras]
  {10.1093/mnras/sty3016}, \href
  {https://ui.adsabs.harvard.edu/\#abs/2019MNRAS.482.4570G} {482, 4570}

\bibitem[\protect\citeauthoryear{{Girven}, {G{\"a}nsicke}, {Steeghs}  \&
  {Koester}}{{Girven} et~al.}{2011}]{girven_gansicke_2011}
{Girven} J.,  {G{\"a}nsicke} B.~T.,  {Steeghs} D.,   {Koester} D.,  2011,
  \mn@doi [\mnras] {10.1111/j.1365-2966.2011.19337.x}, \href
  {http://adsabs.harvard.edu/abs/2011MNRAS.417.1210G} {417, 1210}

\bibitem[\protect\citeauthoryear{Goulding et~al.,}{Goulding
  et~al.}{2012}]{goulding_2012}
Goulding N.~T.,  et~al., 2012, \mn@doi [Monthly Notices of the Royal
  Astronomical Society] {10.1111/j.1365-2966.2012.21932.x}, 427, 3358–3373

\bibitem[\protect\citeauthoryear{{Green}, {Ali}  \& {Napiwotzki}}{{Green}
  et~al.}{2000}]{green_ali_2000}
{Green} P.~J.,  {Ali} B.,   {Napiwotzki} R.,  2000, \mn@doi [\apj]
  {10.1086/309379}, \href
  {https://ui.adsabs.harvard.edu/abs/2000ApJ...540..992G} {540, 992}

\bibitem[\protect\citeauthoryear{{Grether} \& {Lineweaver}}{{Grether} \&
  {Lineweaver}}{2006}]{grether_lineweaver_2006}
{Grether} D.,  {Lineweaver} C.~H.,  2006, \mn@doi [\apj] {10.1086/500161},
  \href {https://ui.adsabs.harvard.edu/abs/2006ApJ...640.1051G} {640, 1051}

\bibitem[\protect\citeauthoryear{{Hartman} \& {Bakos}}{{Hartman} \&
  {Bakos}}{2016}]{hartman2016}
{Hartman} J.~D.,  {Bakos} G.~{\'A}.,  2016, \mn@doi [Astronomy and Computing]
  {10.1016/j.ascom.2016.05.006}, \href
  {https://ui.adsabs.harvard.edu/#abs/2016A&C....17....1H} {17, 1}

\bibitem[\protect\citeauthoryear{{Hirsch} \& {Heber}}{{Hirsch} \&
  {Heber}}{2009}]{hirsch09}
{Hirsch} H.,  {Heber} U.,  2009, in Journal of Physics Conference Series. p.
  012015, \mn@doi{10.1088/1742-6596/172/1/012015}

\bibitem[\protect\citeauthoryear{{Hoard}, {Wachter}, {Sturch}, {Widhalm},
  {Weiler}, {Pretorius}, {Wellhouse}  \& {Gibiansky}}{{Hoard}
  et~al.}{2007}]{hoard_wachter_2007}
{Hoard} D.~W.,  {Wachter} S.,  {Sturch} L.~K.,  {Widhalm} A.~M.,  {Weiler}
  K.~P.,  {Pretorius} M.~L.,  {Wellhouse} J.~W.,   {Gibiansky} M.,  2007,
  \mn@doi [\aj] {10.1086/517878}, \href
  {https://ui.adsabs.harvard.edu/abs/2007AJ....134...26H} {134, 26}

\bibitem[\protect\citeauthoryear{{Holberg} \& {Bergeron}}{{Holberg} \&
  {Bergeron}}{2006}]{holberg06}
{Holberg} J.~B.,  {Bergeron} P.,  2006, \mn@doi [\aj] {10.1086/505938}, \href
  {http://adsabs.harvard.edu/abs/2006AJ....132.1221H} {132, 1221}

\bibitem[\protect\citeauthoryear{Hubeny}{Hubeny}{1988}]{hubeny1988a}
Hubeny I.,  1988, \mn@doi [Comput. Phys. Commun.]
  {10.1016/0010-4655(88)90177-4}, 52, 103

\bibitem[\protect\citeauthoryear{Hubeny \& Lanz}{Hubeny \&
  Lanz}{1995}]{hubeny1995a}
Hubeny I.,  Lanz T.,  1995, \mn@doi [ApJ] {10.1086/175226}, 439, 875

\bibitem[\protect\citeauthoryear{Hubeny \& Lanz}{Hubeny \&
  Lanz}{2011}]{hubeny2011a}
Hubeny I.,  Lanz T.,  2011, in Astrophysics Source Code Library, record
  ascl:1109.022. p.~9022

\bibitem[\protect\citeauthoryear{{Hubeny} \& {Lanz}}{{Hubeny} \&
  {Lanz}}{2017}]{TLUSTY_2017}
{Hubeny} I.,  {Lanz} T.,  2017, arXiv e-prints, \href
  {https://ui.adsabs.harvard.edu/\#abs/2017arXiv170601859H} {p.
  arXiv:1706.01859}

\bibitem[\protect\citeauthoryear{Kepler et~al.,}{Kepler
  et~al.}{2015}]{kepler_2015}
Kepler S.~O.,  et~al., 2015, \mn@doi [Monthly Notices of the Royal Astronomical
  Society] {10.1093/mnras/stu2388}, 446, 4078

\bibitem[\protect\citeauthoryear{Kepler et~al.,}{Kepler
  et~al.}{2016}]{kepler_2016}
Kepler S.~O.,  et~al., 2016, \mn@doi [Monthly Notices of the Royal Astronomical
  Society] {10.1093/mnras/stv2526}, 455, 3413

\bibitem[\protect\citeauthoryear{{Kirkpatrick} et~al.,}{{Kirkpatrick}
  et~al.}{1999a}]{kirkpatrick_davy_1999}
{Kirkpatrick} J.~D.,  et~al., 1999a, \mn@doi [\apj] {10.1086/307414}, \href
  {https://ui.adsabs.harvard.edu/abs/1999ApJ...519..802K} {519, 802}

\bibitem[\protect\citeauthoryear{{Kirkpatrick}, {Allard}, {Bida}, {Zuckerman},
  {Becklin}, {Chabrier}  \& {Baraffe}}{{Kirkpatrick}
  et~al.}{1999b}]{kirkpatrick_allard_1999}
{Kirkpatrick} J.~D.,  {Allard} F.,  {Bida} T.,  {Zuckerman} B.,  {Becklin}
  E.~E.,  {Chabrier} G.,   {Baraffe} I.,  1999b, \mn@doi [\apj]
  {10.1086/307380}, \href
  {https://ui.adsabs.harvard.edu/abs/1999ApJ...519..834K} {519, 834}

\bibitem[\protect\citeauthoryear{{Kleinman} et~al.,}{{Kleinman}
  et~al.}{2013}]{kleinman_kepler_2013}
{Kleinman} S.~J.,  et~al., 2013, \mn@doi [\apjs] {10.1088/0067-0049/204/1/5},
  \href {http://adsabs.harvard.edu/abs/2013ApJS..204....5K} {204, 5}

\bibitem[\protect\citeauthoryear{{Koester} \& {Kepler}}{{Koester} \&
  {Kepler}}{2015}]{koester15}
{Koester} D.,  {Kepler} S.~O.,  2015, \mn@doi [\aap]
  {10.1051/0004-6361/201527169}, \href
  {http://adsabs.harvard.edu/abs/2015A%26A...583A..86K} {583, A86}

\bibitem[\protect\citeauthoryear{{Lawrence} et~al.,}{{Lawrence}
  et~al.}{2007}]{ukidds}
{Lawrence} A.,  et~al., 2007, \mn@doi [\mnras]
  {10.1111/j.1365-2966.2007.12040.x}, \href
  {http://adsabs.harvard.edu/abs/2007MNRAS.379.1599L} {379, 1599}

\bibitem[\protect\citeauthoryear{{Lawrie}, {Burleigh}, {Dufour}  \&
  {Hodgkin}}{{Lawrie} et~al.}{2013}]{lawrie13_1}
{Lawrie} K.~A.,  {Burleigh} M.~R.,  {Dufour} P.,   {Hodgkin} S.~T.,  2013,
  \mn@doi [\mnras] {10.1093/mnras/stt832}, \href
  {http://adsabs.harvard.edu/abs/2013MNRAS.433.1599L} {433, 1599}

\bibitem[\protect\citeauthoryear{{Lee}, {Berger}  \& {Knapp}}{{Lee}
  et~al.}{2010}]{lee_berger_2010}
{Lee} K.-G.,  {Berger} E.,   {Knapp} G.~R.,  2010, \mn@doi [\apj]
  {10.1088/0004-637X/708/2/1482}, \href
  {https://ui.adsabs.harvard.edu/abs/2010ApJ...708.1482L} {708, 1482}

\bibitem[\protect\citeauthoryear{{Lenz} \& {Breger}}{{Lenz} \&
  {Breger}}{2005}]{lenz2005}
{Lenz} P.,  {Breger} M.,  2005, \mn@doi [Communications in Asteroseismology]
  {10.1553/cia146s53}, \href
  {https://ui.adsabs.harvard.edu/#abs/2005CoAst.146...53L} {146, 53}

\bibitem[\protect\citeauthoryear{{Liebert}, {Bergeron}  \& {Holberg}}{{Liebert}
  et~al.}{2005}]{liebert_bergeron_2005}
{Liebert} J.,  {Bergeron} P.,   {Holberg} J.~B.,  2005, \mn@doi [\apjs]
  {10.1086/425738}, \href {http://adsabs.harvard.edu/abs/2005ApJS..156...47L}
  {156, 47}

\bibitem[\protect\citeauthoryear{{Lomb}}{{Lomb}}{1976}]{Lomb1976}
{Lomb} N.~R.,  1976, \mn@doi [\apss] {10.1007/BF00648343}, \href
  {https://ui.adsabs.harvard.edu/#abs/1976Ap&SS..39..447L} {39, 447}

\bibitem[\protect\citeauthoryear{{Longstaff}, {Casewell}, {Wynn}, {Maxted}  \&
  {Helling}}{{Longstaff} et~al.}{2017}]{longstaff_casewell_2017}
{Longstaff} E.~S.,  {Casewell} S.~L.,  {Wynn} G.~A.,  {Maxted} P.~F.~L.,
  {Helling} C.,  2017, \mn@doi [\mnras] {10.1093/mnras/stx1786}, \href
  {http://adsabs.harvard.edu/abs/2017MNRAS.471.1728L} {471, 1728}

\bibitem[\protect\citeauthoryear{{Longstaff}, {Casewell}, {Wynn}, {Page},
  {Williams}, {Braker}  \& {Maxted}}{{Longstaff}
  et~al.}{2019}]{longstaff_casewell_2019}
{Longstaff} E.~S.,  {Casewell} S.~L.,  {Wynn} G.~A.,  {Page} K.~L.,  {Williams}
  P.~K.~G.,  {Braker} I.,   {Maxted} P.~F.~L.,  2019, \mn@doi [\mnras]
  {10.1093/mnras/stz127}, \href
  {http://adsabs.harvard.edu/abs/2019MNRAS.484.2566L} {484, 2566}

\bibitem[\protect\citeauthoryear{Luo, Németh, Liu, Deng  \& Han}{Luo
  et~al.}{2016}]{luo_nemeth_2016}
Luo Y.-P.,  Németh P.,  Liu C.,  Deng L.-C.,   Han Z.-W.,  2016, The
  Astrophysical Journal, 818, 202

\bibitem[\protect\citeauthoryear{Marcy \& Butler}{Marcy \&
  Butler}{2000}]{marcy_butler_2000}
Marcy G.,  Butler R.,  2000, Publications of the Astronomical Society of the
  Pacific, 112, 137

\bibitem[\protect\citeauthoryear{{Marsh}, {Dhillon}  \& {Duck}}{{Marsh}
  et~al.}{1995}]{marsh95}
{Marsh} T.~R.,  {Dhillon} V.~S.,   {Duck} S.~R.,  1995, \mn@doi [\mnras]
  {10.1093/mnras/275.3.828}, \href
  {https://ui.adsabs.harvard.edu/abs/1995MNRAS.275..828M} {275, 828}

\bibitem[\protect\citeauthoryear{{Maxted}, {Napiwotzki}, {Dobbie}  \&
  {Burleigh}}{{Maxted} et~al.}{2006}]{maxted_napiwotzki_2006}
{Maxted} P.~F.~L.,  {Napiwotzki} R.,  {Dobbie} P.~D.,   {Burleigh} M.~R.,
  2006, \mn@doi [\nat] {10.1038/nature04987}, \href
  {http://adsabs.harvard.edu/abs/2006Natur.442..543M} {442, 543}

\bibitem[\protect\citeauthoryear{{Metchev} \& {Hillenbrand}}{{Metchev} \&
  {Hillenbrand}}{2009}]{metchev_hillenbrand_2009}
{Metchev} S.~A.,  {Hillenbrand} L.~A.,  2009, \mn@doi [\apjs]
  {10.1088/0067-0049/181/1/62}, \href
  {http://adsabs.harvard.edu/abs/2009ApJS..181...62M} {181, 62}

\bibitem[\protect\citeauthoryear{{Napiwotzki}, {Green}  \&
  {Saffer}}{{Napiwotzki} et~al.}{1999}]{napiwotzki_green_saffer}
{Napiwotzki} R.,  {Green} P.~J.,   {Saffer} R.~A.,  1999, \mn@doi [\apj]
  {10.1086/307170}, \href {http://adsabs.harvard.edu/abs/1999ApJ...517..399N}
  {517, 399}

\bibitem[\protect\citeauthoryear{{Napiwotzki} et~al.,}{{Napiwotzki}
  et~al.}{2004}]{napiwotzki_yungelson_2004}
{Napiwotzki} R.,  et~al., 2004, in {Hilditch} R.~W.,  {Hensberge} H.,
  {Pavlovski} K.,  eds,  Astronomical Society of the Pacific Conference Series
  Vol. 318, Spectroscopically and Spatially Resolving the Components of the
  Close Binary Stars. pp 402--410

\bibitem[\protect\citeauthoryear{{Naylor}}{{Naylor}}{1998}]{naylor_1998}
{Naylor} T.,  1998, \mn@doi [\mnras] {10.1046/j.1365-8711.1998.01314.x}, \href
  {https://ui.adsabs.harvard.edu/abs/1998MNRAS.296..339N} {296, 339}

\bibitem[\protect\citeauthoryear{{O'Toole} \& {Heber}}{{O'Toole} \&
  {Heber}}{2006}]{otoole06}
{O'Toole} S.~J.,  {Heber} U.,  2006, \mn@doi [\aap]
  {10.1051/0004-6361:20053948}, \href
  {http://adsabs.harvard.edu/abs/2006A%26A...452..579O} {452, 579}

\bibitem[\protect\citeauthoryear{{Parsons} et~al.,}{{Parsons}
  et~al.}{2017}]{parsons_hermes_2017}
{Parsons} S.~G.,  et~al., 2017, \mn@doi [\mnras] {10.1093/mnras/stx1610}, \href
  {http://adsabs.harvard.edu/abs/2017MNRAS.471..976P} {471, 976}

\bibitem[\protect\citeauthoryear{{Probst}}{{Probst}}{1983}]{probst_1983}
{Probst} R.~G.,  1983, \mn@doi [\apj] {10.1086/161441}, \href
  {https://ui.adsabs.harvard.edu/abs/1983ApJ...274..237P} {274, 237}

\bibitem[\protect\citeauthoryear{{Probst} \& {O'Connell}}{{Probst} \&
  {O'Connell}}{1982}]{probst_oconnell_1982}
{Probst} R.~G.,  {O'Connell} R.~W.,  1982, \mn@doi [\apjl] {10.1086/183722},
  \href {https://ui.adsabs.harvard.edu/abs/1982ApJ...252L..69P} {252, L69}

\bibitem[\protect\citeauthoryear{Radigan, Lafrenière, Jayawardhana  \&
  Artigau}{Radigan et~al.}{2014}]{radigan_2014}
Radigan J.,  Lafrenière D.,  Jayawardhana R.,   Artigau E.,  2014, \mn@doi
  [The Astrophysical Journal] {10.1088/0004-637x/793/2/75}, 793, 75

\bibitem[\protect\citeauthoryear{{Rayner}, {Cushing}  \& {Vacca}}{{Rayner}
  et~al.}{2009}]{rayner_cushing_2009}
{Rayner} J.~T.,  {Cushing} M.~C.,   {Vacca} W.~D.,  2009, \mn@doi [\apjs]
  {10.1088/0067-0049/185/2/289}, \href
  {http://adsabs.harvard.edu/abs/2009ApJS..185..289R} {185, 289}

\bibitem[\protect\citeauthoryear{Rebassa-Mansergas, Agurto-Gangas, Schreiber,
  Gänsicke  \& Koester}{Rebassa-Mansergas
  et~al.}{2013}]{rabassa_mansegas_2013}
Rebassa-Mansergas A.,  Agurto-Gangas C.,  Schreiber M.~R.,  Gänsicke B.~T.,
  Koester D.,  2013, \mn@doi [Monthly Notices of the Royal Astronomical
  Society] {10.1093/mnras/stt974}, 433, 3398

\bibitem[\protect\citeauthoryear{{Sahlmann} et~al.,}{{Sahlmann}
  et~al.}{2011}]{sahlmann_segransan_2011}
{Sahlmann} J.,  et~al., 2011, \mn@doi [\aap] {10.1051/0004-6361/201015427},
  \href {https://ui.adsabs.harvard.edu/abs/2011A&A...525A..95S} {525, A95}

\bibitem[\protect\citeauthoryear{{Scargle}}{{Scargle}}{1982}]{Scargle1982}
{Scargle} J.~D.,  1982, \mn@doi [\apj] {10.1086/160554}, \href
  {https://ui.adsabs.harvard.edu/#abs/1982ApJ...263..835S} {263, 835}

\bibitem[\protect\citeauthoryear{{Schaffenroth}, {Barlow}, {Drechsel}  \&
  {Dunlap}}{{Schaffenroth} et~al.}{2015}]{schaffenroth}
{Schaffenroth} V.,  {Barlow} B.~N.,  {Drechsel} H.,   {Dunlap} B.~H.,  2015,
  \mn@doi [\aap] {10.1051/0004-6361/201525701}, \href
  {http://adsabs.harvard.edu/abs/2015A%26A...576A.123S} {576, A123}

\bibitem[\protect\citeauthoryear{{Steele}, {Burleigh}, {Dobbie}, {Jameson},
  {Barstow}  \& {Satterthwaite}}{{Steele} et~al.}{2011}]{steele_burleigh_2011}
{Steele} P.~R.,  {Burleigh} M.~R.,  {Dobbie} P.~D.,  {Jameson} R.~F.,
  {Barstow} M.~A.,   {Satterthwaite} R.~P.,  2011, \mn@doi [\mnras]
  {10.1111/j.1365-2966.2011.19225.x}, \href
  {http://adsabs.harvard.edu/abs/2011MNRAS.416.2768S} {416, 2768}

\bibitem[\protect\citeauthoryear{{Steele} et~al.,}{{Steele}
  et~al.}{2013a}]{steele_saglia_2013}
{Steele} P.~R.,  et~al., 2013a, \mn@doi [\mnras] {10.1093/mnras/sts620}, \href
  {http://adsabs.harvard.edu/abs/2013MNRAS.429.3492S} {429, 3492}

\bibitem[\protect\citeauthoryear{Steele et~al.,}{Steele
  et~al.}{2013b}]{steele_sagalia_2013}
Steele P.~R.,  et~al., 2013b, \mn@doi [Monthly Notices of the Royal
  Astronomical Society] {10.1093/mnras/sts620}, 429, 3492

\bibitem[\protect\citeauthoryear{{Stroeer}, {Heber}, {Lisker}, {Napiwotzki},
  {Dreizler}, {Christlieb}  \& {Reimers}}{{Stroeer} et~al.}{2007}]{stro}
{Stroeer} A.,  {Heber} U.,  {Lisker} T.,  {Napiwotzki} R.,  {Dreizler} S.,
  {Christlieb} N.,   {Reimers} D.,  2007, \mn@doi [\aap]
  {10.1051/0004-6361:20065564}, \href
  {http://adsabs.harvard.edu/abs/2007A%26A...462..269S} {462, 269}

\bibitem[\protect\citeauthoryear{{Tan} \& {Showman}}{{Tan} \&
  {Showman}}{2020}]{tan_showman_2020}
{Tan} X.,  {Showman} A.~P.,  2020, arXiv e-prints, \href
  {https://ui.adsabs.harvard.edu/abs/2020arXiv200106269T} {p. arXiv:2001.06269}

\bibitem[\protect\citeauthoryear{{Tody}}{{Tody}}{1993}]{tody_1993}
{Tody} D.,  1993, in {Hanisch} R.~J.,  {Brissenden} R.~J.~V.,   {Barnes} J.,
  eds,  Astronomical Society of the Pacific Conference Series Vol. 52,
  Astronomical Data Analysis Software and Systems II. p.~173

\bibitem[\protect\citeauthoryear{Tremblay \& Bergeron}{Tremblay \&
  Bergeron}{2009}]{tremblay2009a}
Tremblay P.~E.,  Bergeron P.,  2009, \mn@doi [ApJ]
  {10.1088/0004-637X/696/2/1755}, 696, 1755

\bibitem[\protect\citeauthoryear{{Tremblay}, {Ludwig}, {Steffen}, {Bergeron}
  \& {Freytag}}{{Tremblay} et~al.}{2011a}]{tremblay_2011}
{Tremblay} P.~E.,  {Ludwig} H.~G.,  {Steffen} M.,  {Bergeron} P.,   {Freytag}
  B.,  2011a, \mn@doi [\aap] {10.1051/0004-6361/201117310}, \href
  {https://ui.adsabs.harvard.edu/abs/2011A&A...531L..19T} {531, L19}

\bibitem[\protect\citeauthoryear{{Tremblay}, {Bergeron}  \&
  {Gianninas}}{{Tremblay} et~al.}{2011b}]{tremblay_bergeron_2011}
{Tremblay} P.~E.,  {Bergeron} P.,   {Gianninas} A.,  2011b, \mn@doi [\apj]
  {10.1088/0004-637X/730/2/128}, \href
  {https://ui.adsabs.harvard.edu/abs/2011ApJ...730..128T} {730, 128}

\bibitem[\protect\citeauthoryear{{Tremblay}, {Cukanovaite}, {Gentile Fusillo},
  {Cunningham}  \& {Hollands}}{{Tremblay} et~al.}{2019}]{tremblay19}
{Tremblay} P.~E.,  {Cukanovaite} E.,  {Gentile Fusillo} N.~P.,  {Cunningham}
  T.,   {Hollands} M.~A.,  2019, \mn@doi [\mnras] {10.1093/mnras/sty3067},
  \href {https://ui.adsabs.harvard.edu/abs/2019MNRAS.482.5222T} {482, 5222}

\bibitem[\protect\citeauthoryear{{Triaud} et~al.,}{{Triaud}
  et~al.}{2017}]{triaud_amaury_2017}
{Triaud} A. H.~M.~J.,  et~al., 2017, \mn@doi [\aap]
  {10.1051/0004-6361/201730993}, \href
  {https://ui.adsabs.harvard.edu/abs/2017A&A...608A.129T} {608, A129}

\bibitem[\protect\citeauthoryear{{Troup} et~al.,}{{Troup}
  et~al.}{2016}]{troup_nidever_2016}
{Troup} N.~W.,  et~al., 2016, \mn@doi [\aj] {10.3847/0004-6256/151/3/85}, \href
  {http://adsabs.harvard.edu/abs/2016AJ....151...85T} {151, 85}

\bibitem[\protect\citeauthoryear{{Vacca}, {Cushing}  \& {Rayner}}{{Vacca}
  et~al.}{2003}]{vacca03}
{Vacca} W.~D.,  {Cushing} M.~C.,   {Rayner} J.~T.,  2003, \mn@doi [\pasp]
  {10.1086/346193}, \href {http://adsabs.harvard.edu/abs/2003PASP..115..389V}
  {115, 389}

\bibitem[\protect\citeauthoryear{Verbeek et~al.,}{Verbeek
  et~al.}{2014}]{verbeek_2013}
Verbeek K.,  et~al., 2014, \mn@doi [Monthly Notices of the Royal Astronomical
  Society] {10.1093/mnras/stt1492}, 438, 2

\bibitem[\protect\citeauthoryear{Vos et~al.,}{Vos
  et~al.}{2018}]{vos_biller_2018}
Vos J.~M.,  et~al., 2018, \mn@doi [Monthly Notices of the Royal Astronomical
  Society] {10.1093/mnras/sty3123}, 483, 480

\makeatother
\end{thebibliography}
%\newpage

%%%%%%%%%%%%%%%%%%%%%%%%%%%%%%%%%%%%%% APPENDIX %%%%%%%%%%%%
\appendix
%\clearpage
\section{Balmer Line Fits}
We present here the figures associated with the effective temperature and log $g$ fitting to the Balmer lines.
In each figure the data is shown with the black solid line and the model with the red solid line. The dotted line shows features that were removed from the fit.
%\newpage
%%%%%%%%%%%%%%%%%BALMER LINE FIGURES HERE %%%%%%%%%%%%%%%%%%%%%%

%\begin{minipage}{\columnwidth}
%\begin{center}
%\includegraphics[width=0.9\columnwidth]{OB14_BalmerLines.pdf}
%\captionof{figure}{Balmer Line fit of SDSS J010405.12$+$145907.2}
%\label{fig:OB14BL}
%\end{center}
%\end{minipage}

\begin{figure}
\includegraphics[width=0.9\columnwidth]{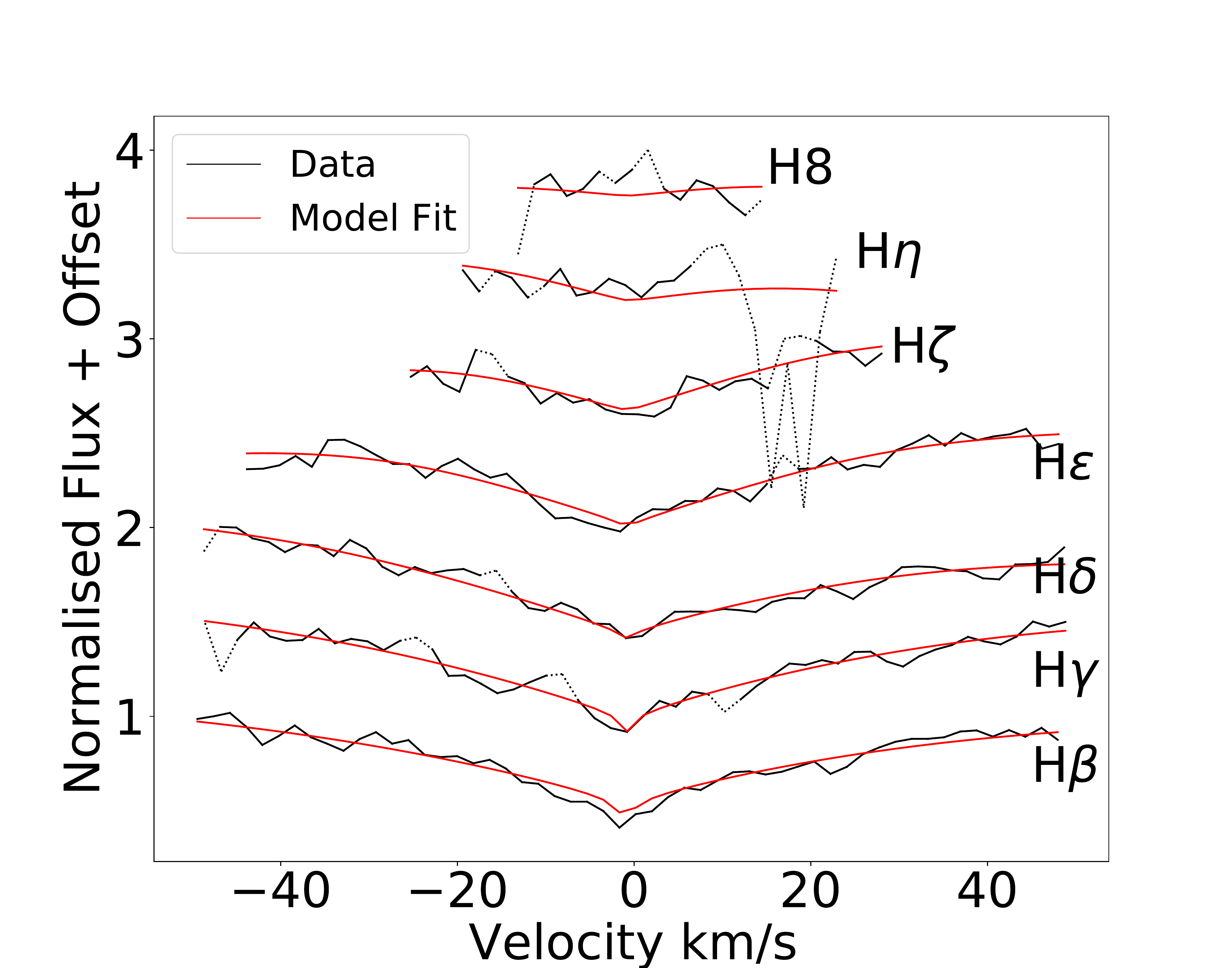}
\caption{Balmer Line fit of SDSS J010405.12$+$145907.2. The black solid line is the data and the model is shown with the red solid line. The dotted line shows features that were removed from the fit.}
\label{fig:OB14BL}
\end{figure}

\begin{figure}
\includegraphics[width=0.9\columnwidth]{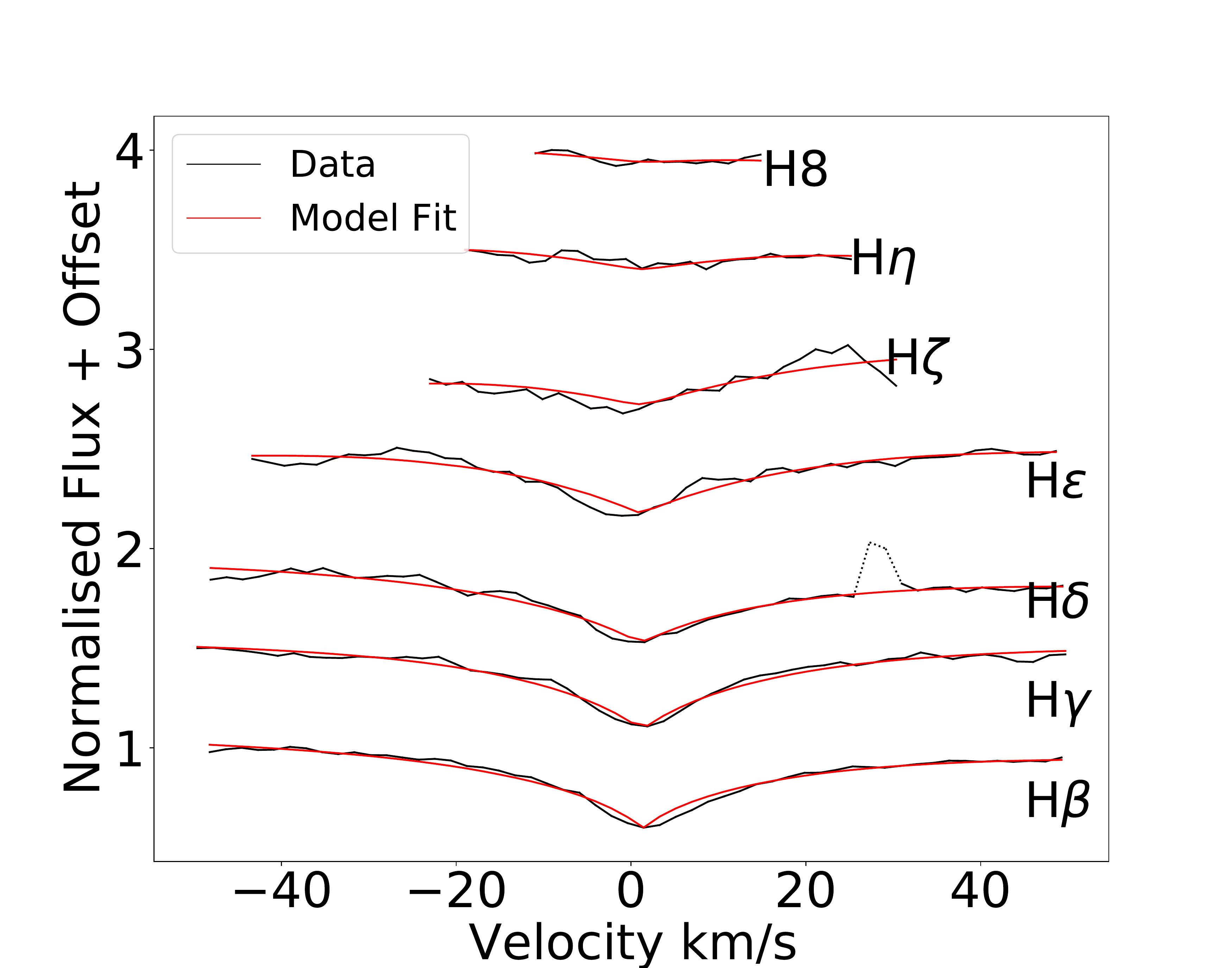}
\caption{Balmer Line fit of SDSS\,J074231.98$+$285727.3. The black solid line is the data and the model is shown with the red solid line. The dotted line shows features that were removed from the fit.}
\label{fig:0742BL}
\end{figure}

\begin{figure}
\includegraphics[width=0.9\columnwidth]{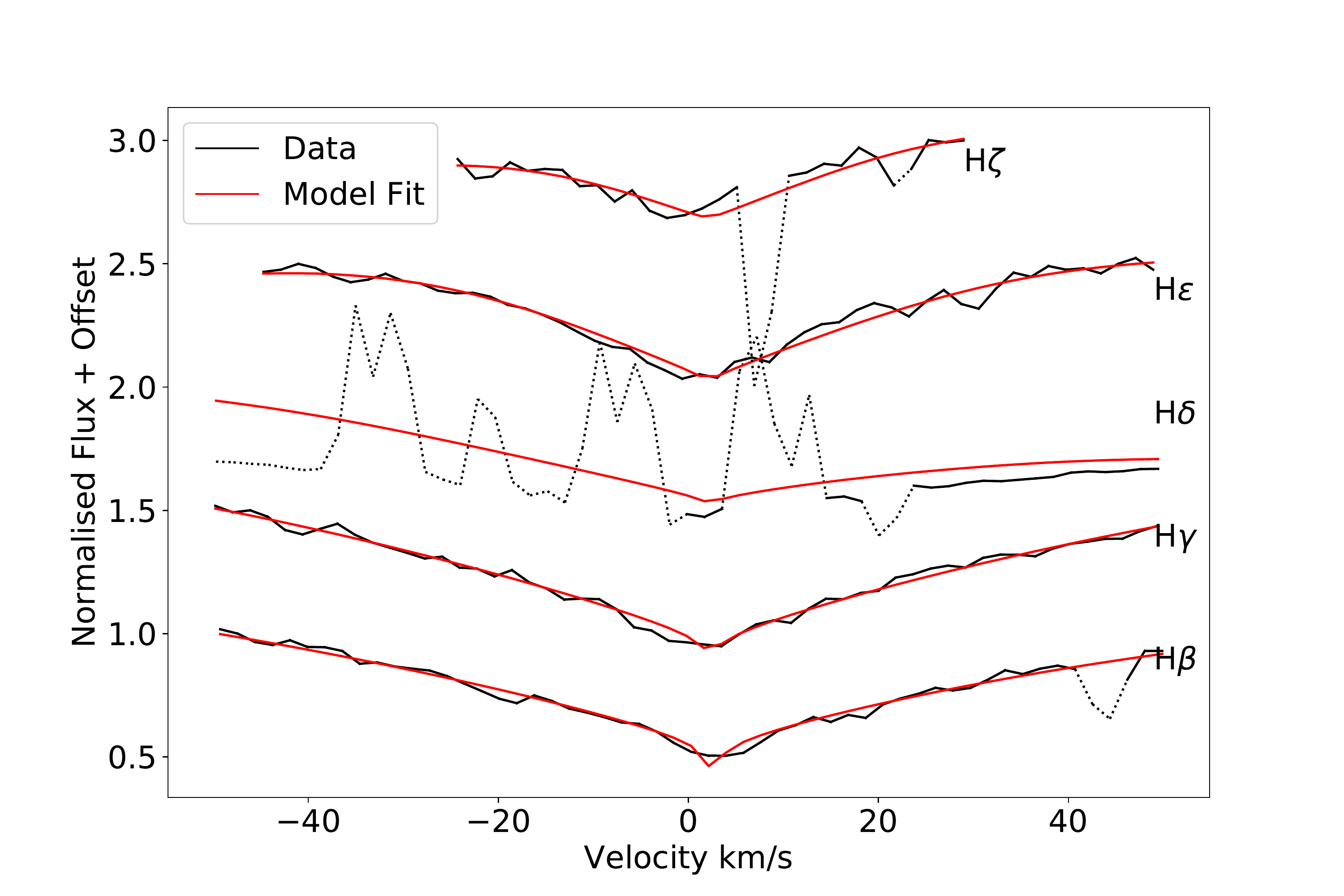}
\caption{Balmer line fit of SDSS J092534.99$-$014046.8. The black solid line is the data and the model is shown with the red solid line. The dotted line shows features that were removed from the fit.}
\label{fig:OB1BL}
\end{figure} 

\begin{figure}
\includegraphics[width=0.9\columnwidth]{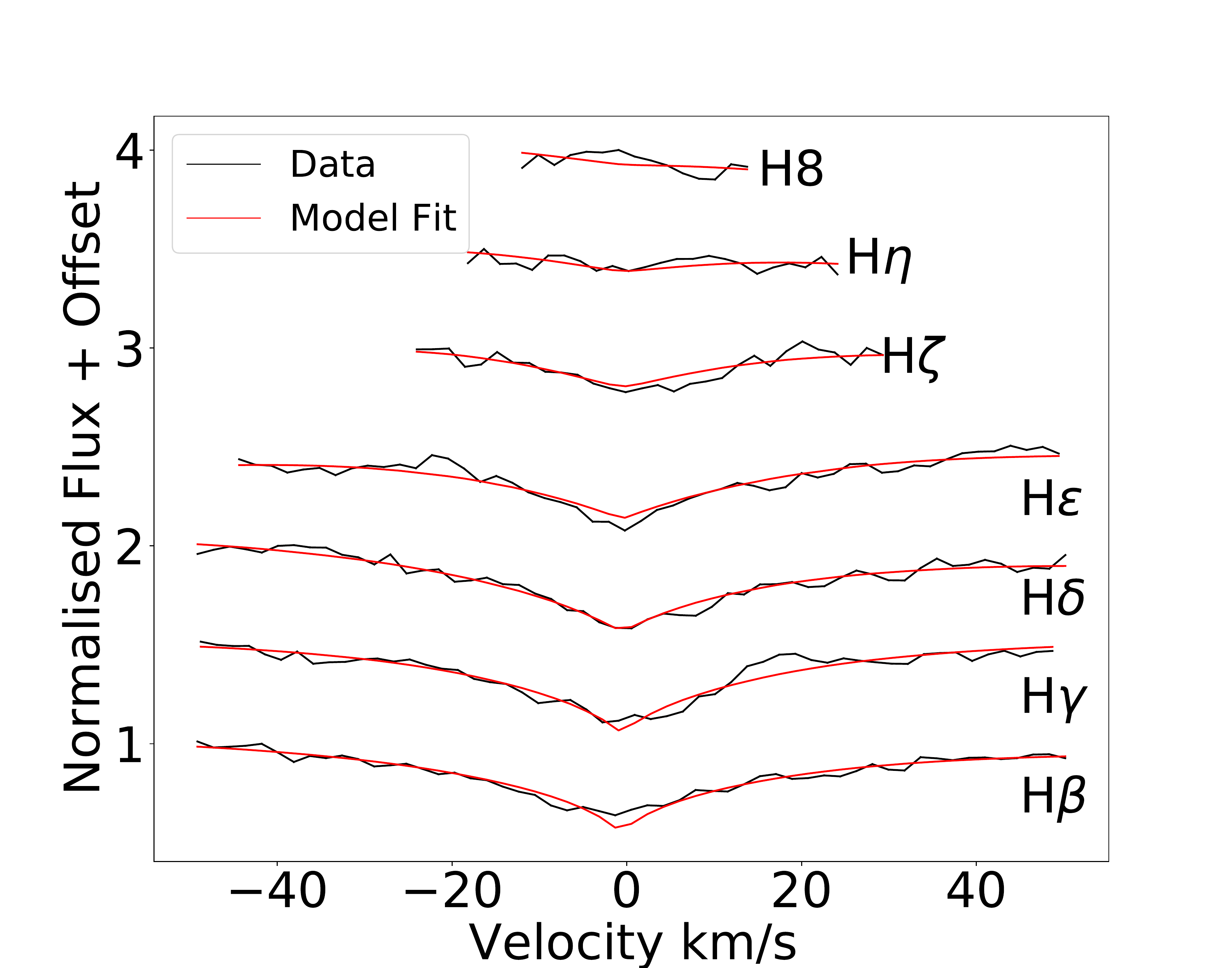}
\caption{Balmer line fit of SDSS\,J103844.58$+$110053.5. The black solid line is the data and the model is shown with the red solid line.}
\label{fig:1038BL}
\end{figure} 

\begin{figure}
\includegraphics[width=0.9\columnwidth]{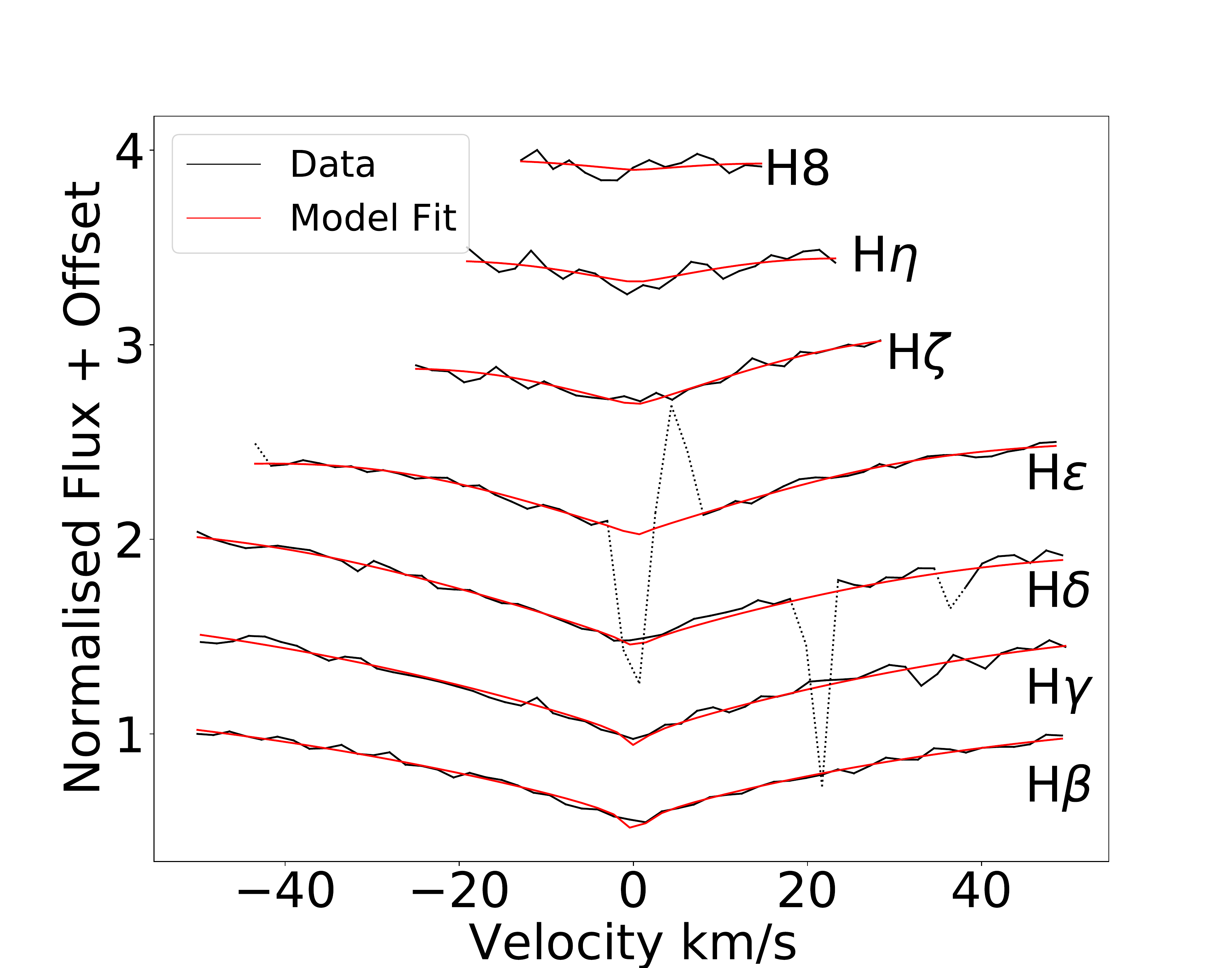}
\caption{Balmer line fit of SDSS\,J154806.89$+$000639.4. The black solid line is the data and the model is shown with the red solid line. The dotted line shows features that were removed from the fit.}
\label{fig:OB3BL}
\end{figure}

\begin{figure}
\includegraphics[width=0.9\columnwidth]{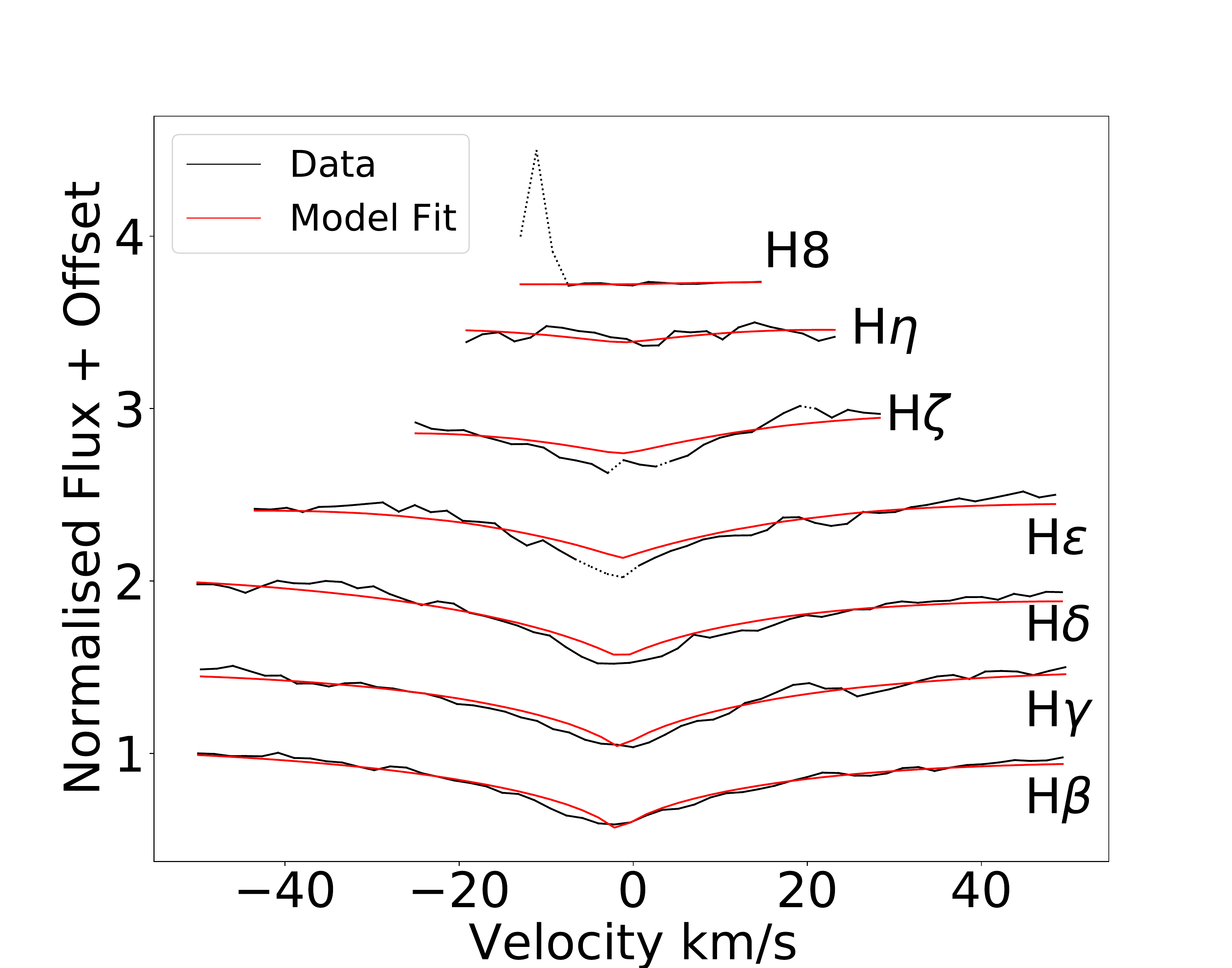}
\caption{Balmer line fit of UVEX\,J184610.80$+$022032.4. The black solid line is the data and the model is shown with the red solid line. The dotted line shows features that were removed from the fit.}
\label{fig:1846BL}
\end{figure}

\begin{figure}
\includegraphics[width=0.9\columnwidth]{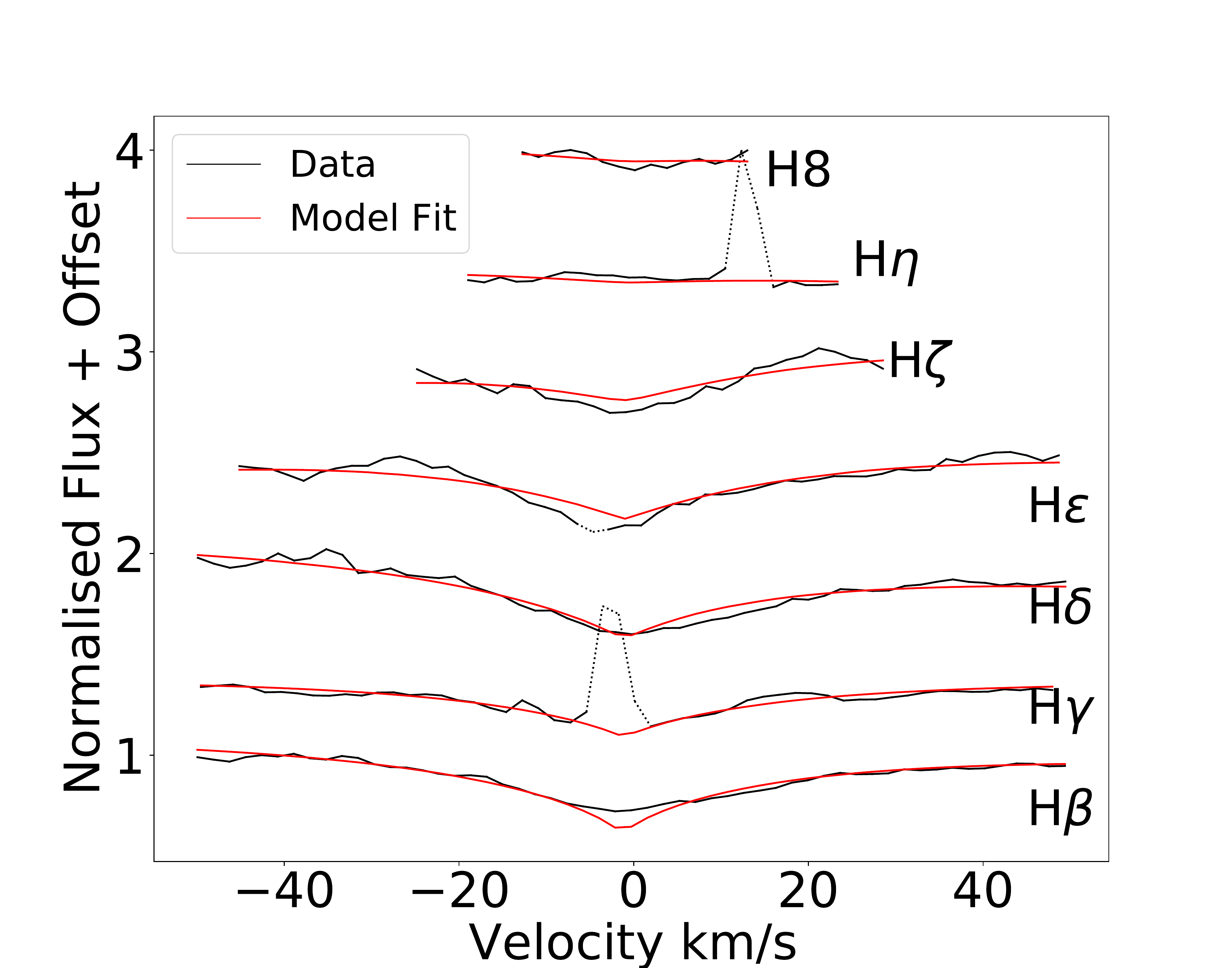}
\caption{Balmer line fit of UVEX\,J185941.43$+$013954.0. The black solid line is the data and the model is shown with the red solid line. The dotted line shows features that were removed from the fit.}
\label{fig:1859BL}
\end{figure}

\begin{figure}
\includegraphics[width=0.9\columnwidth]{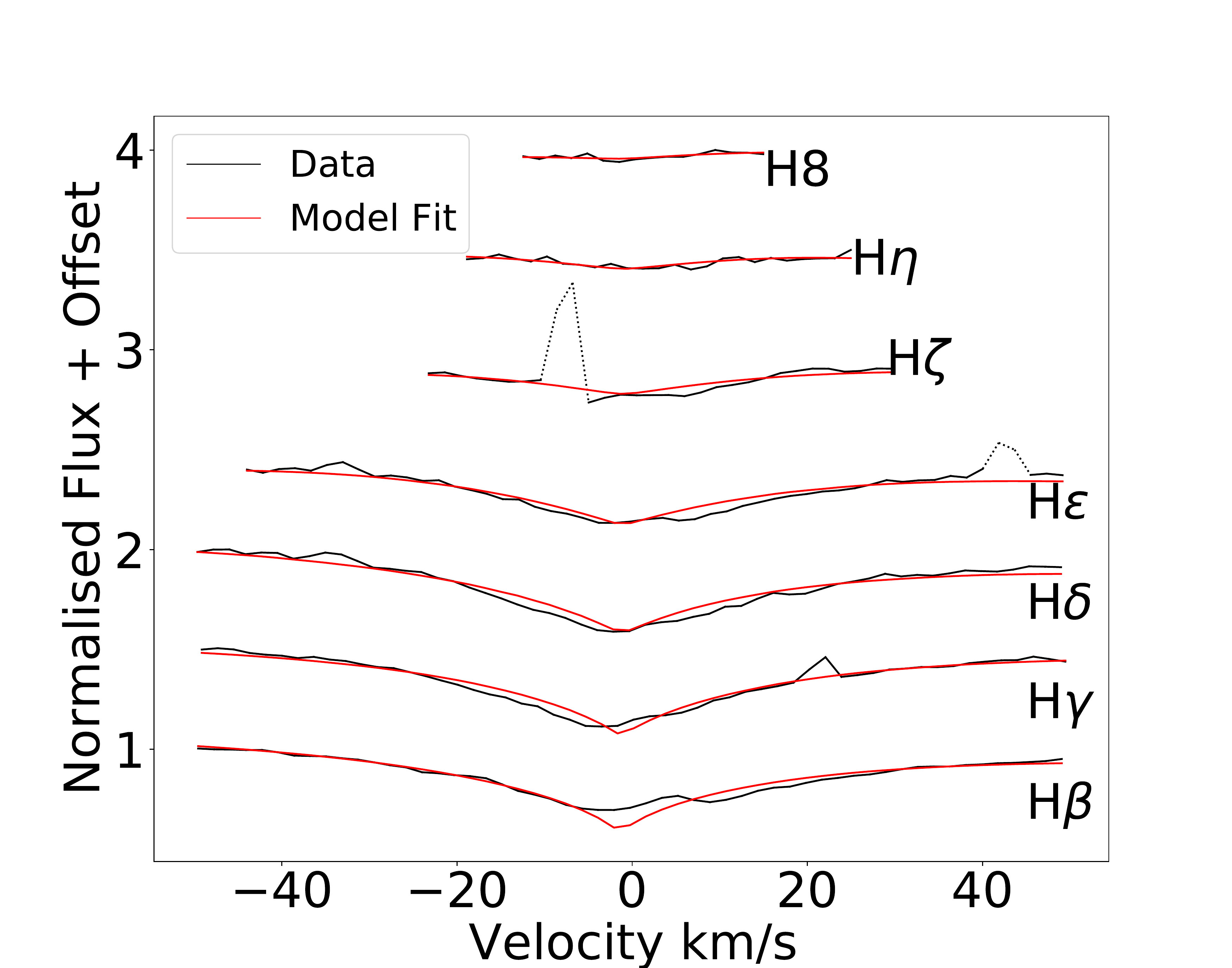}
\caption{Balmer line fit of 2MASS\,J20265915$+$4116436. The black solid line is the data and the model is shown with the red solid line. The dotted line shows features that were removed from the fit.}
\label{fig:2026BL}
\end{figure}

\begin{figure}
\includegraphics[width=0.9\columnwidth]{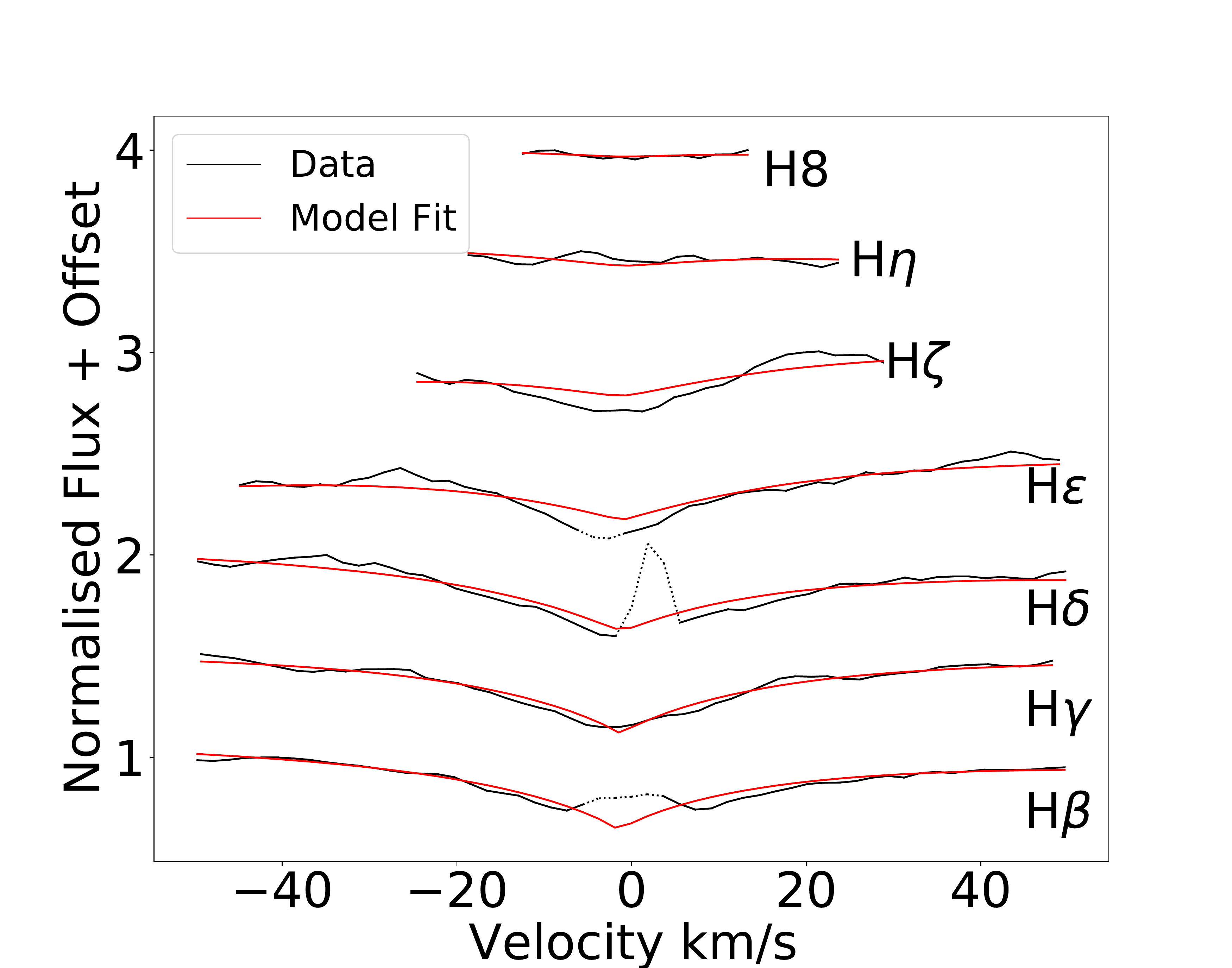}
\caption{Balmer line fit of UVEX\,J204229.67$+$384058.0. The black solid line is the data and the model is shown with the red solid line. The dotted line shows features that were removed from the fit.}
\label{fig:2042BL}
\end{figure}
\clearpage

\begin{figure}
\includegraphics[width=0.9\columnwidth]{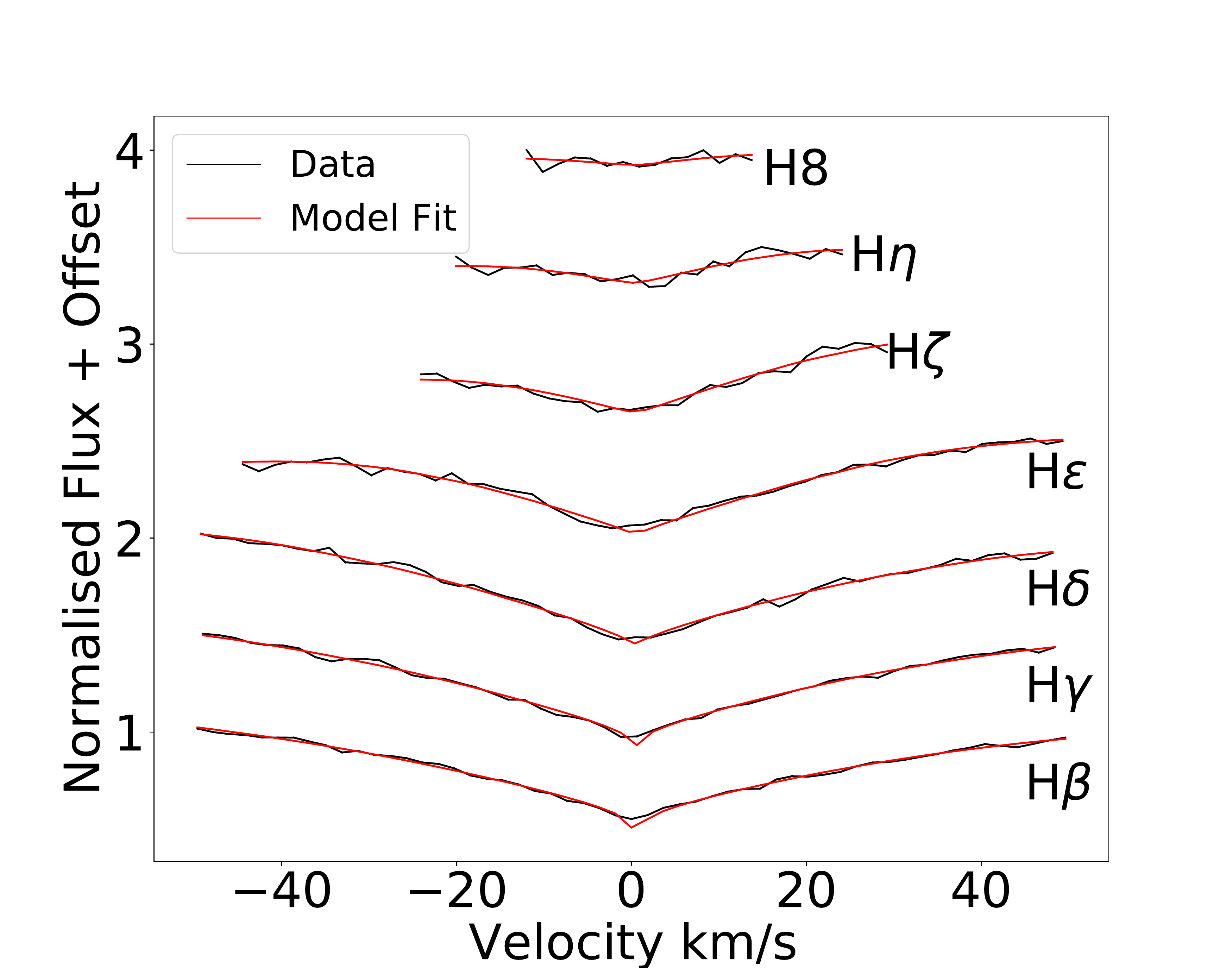}
\caption{Balmer line fit of UGPS\,J210248.46$+$475058.6. The black solid line is the data and the model is shown with the red solid line.}
\label{fig:2102BL}
\end{figure}

\section{SED Fits}
%\newpage 
We present here the results of the SED and spectral template fitting that were not in the main text of this paper. Each figure includes residuals for the fit to a WD model alone (black squares), and the best fitting combined WD-L or M dwarf template spectrum (green squares).

\begin{figure}
\includegraphics[width=0.9\columnwidth]{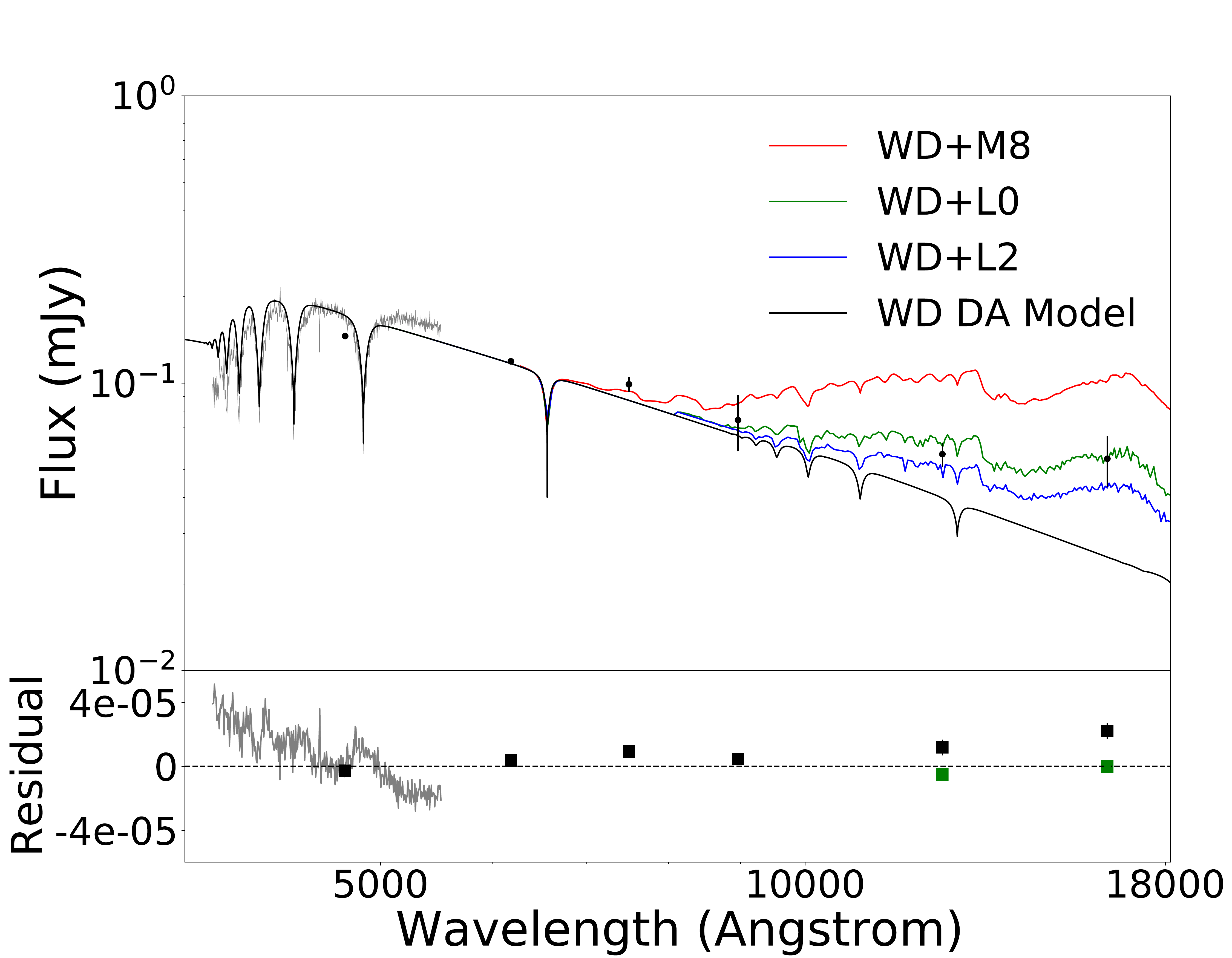}
\caption{OSIRIS spectrum of SDSS J010405.12$+$145907.2, the combined  white dwarf model and brown dwarf template spectra of M8, L0, L2 and the optical and NIR photometry. The data suggest a companion of spectral type between L0-L2}
\label{fig:OB14}
\end{figure}

%\begin{figure}
%\includegraphics[width=0.9\columnwidth]{0742.pdf}
%\caption{OSIRIS spectrum of SDSS074231.98$+$285727.3 with the photometry and the combined spectra of the white dwarf model and M4, M5 and M6 stars for comparison. The data suggest that the companion is between M5-M6}
%\label{fig:0742}
%\end{figure} 

\begin{figure}
\includegraphics[width=0.9\columnwidth]{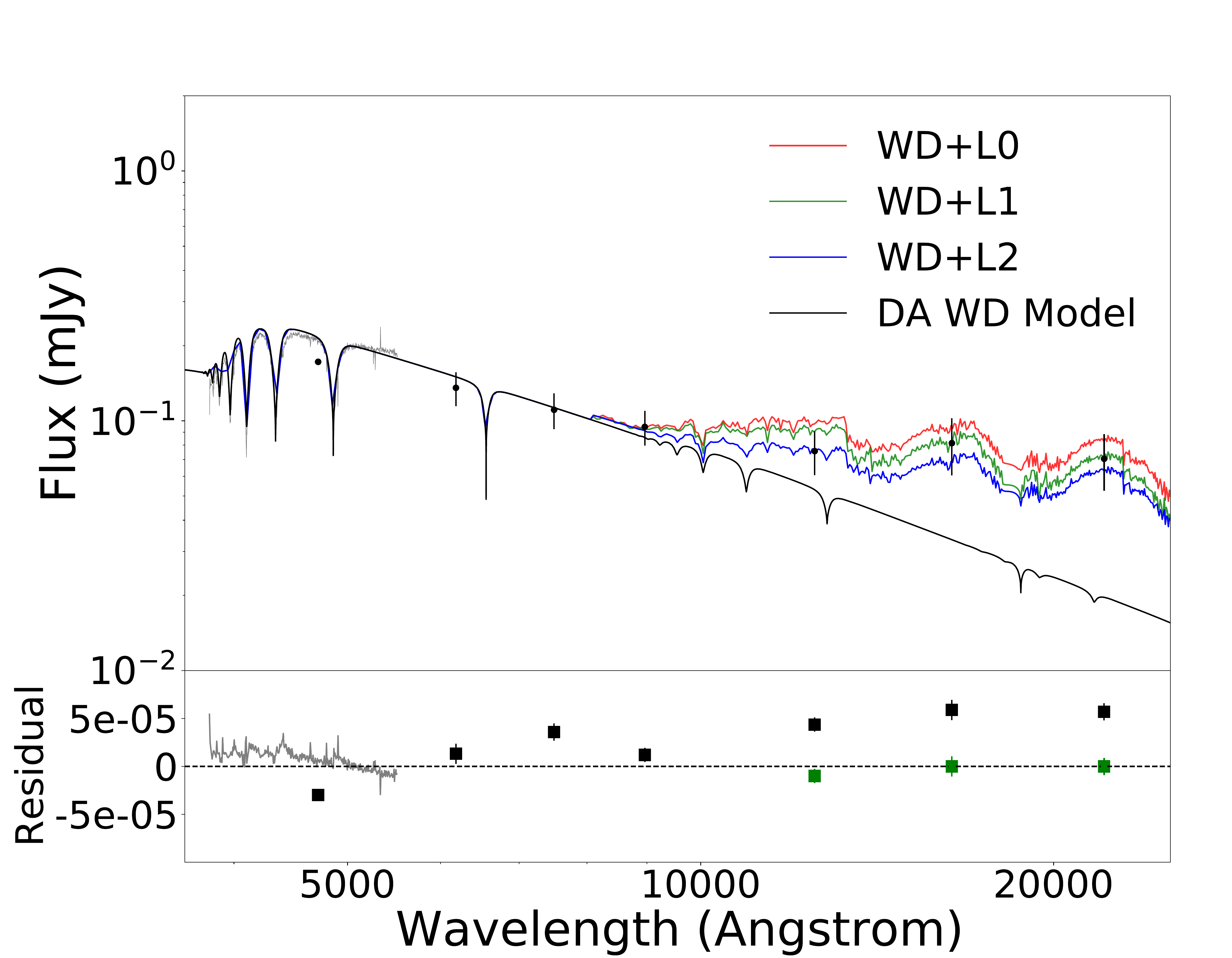}
\caption{OSIRIS spectrum of SDSS J092534.99$-$014046.8 shown with the broadband photometry and the combined DA white dwarf + L0, L1, and L2 brown dwarf template spectra for comparison. The companion is likely to have a spectral type of L1-L2.}
\label{fig:OB1}
\end{figure}

\begin{figure}
\includegraphics[width=0.9\columnwidth, angle=0]{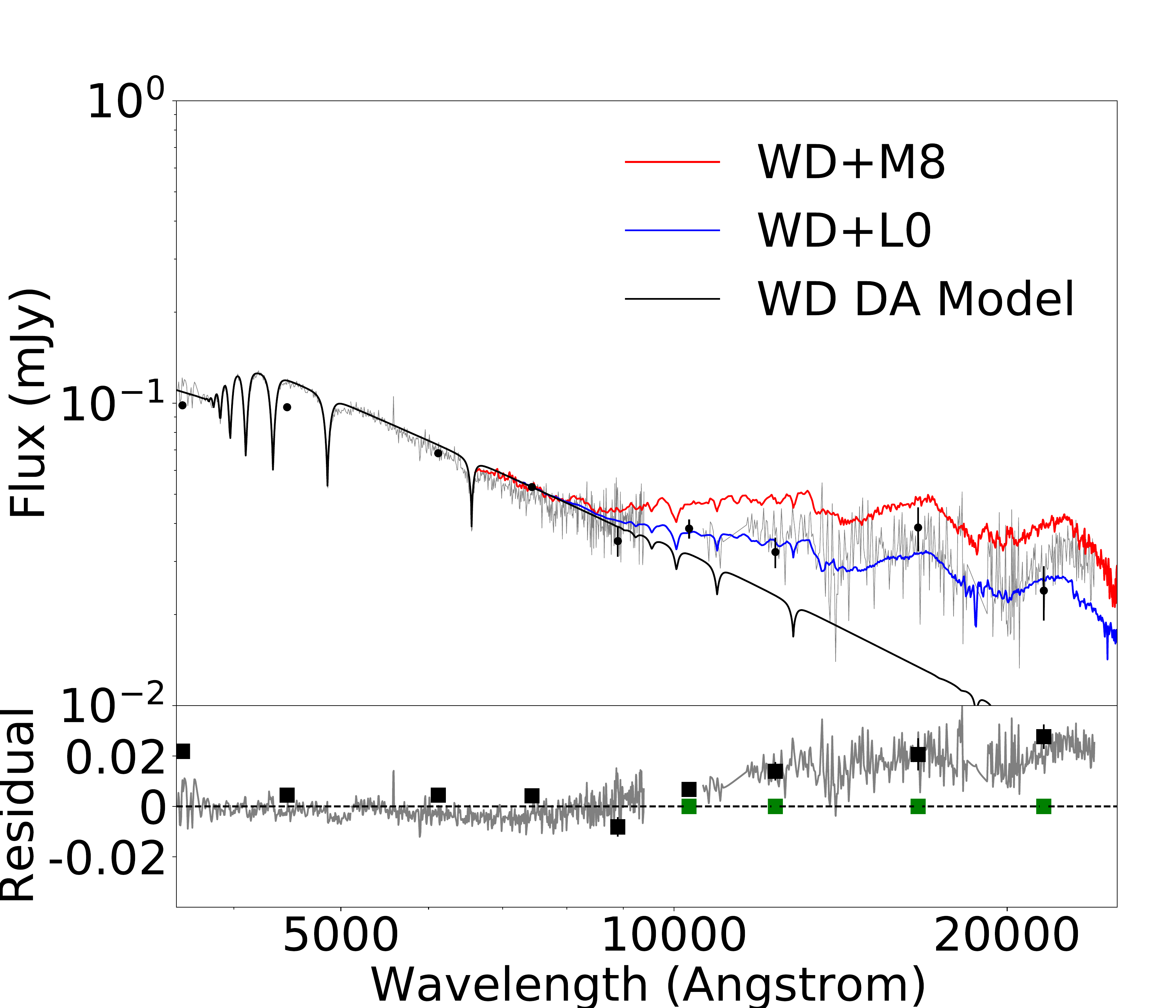}
\caption{Optical (SDSS) and GNIRS NIR spectra of SDSS100300.08$+$093940.16 (S/N$\sim$5) shown with the DA white dwarf model, and combined white dwarf+M8 and white dwarf+L0 models. The SDSS and UKIDSS photometry is also shown.}
\label{1003}
\end{figure}

\begin{figure}
\includegraphics[width=0.9\columnwidth, angle=0]{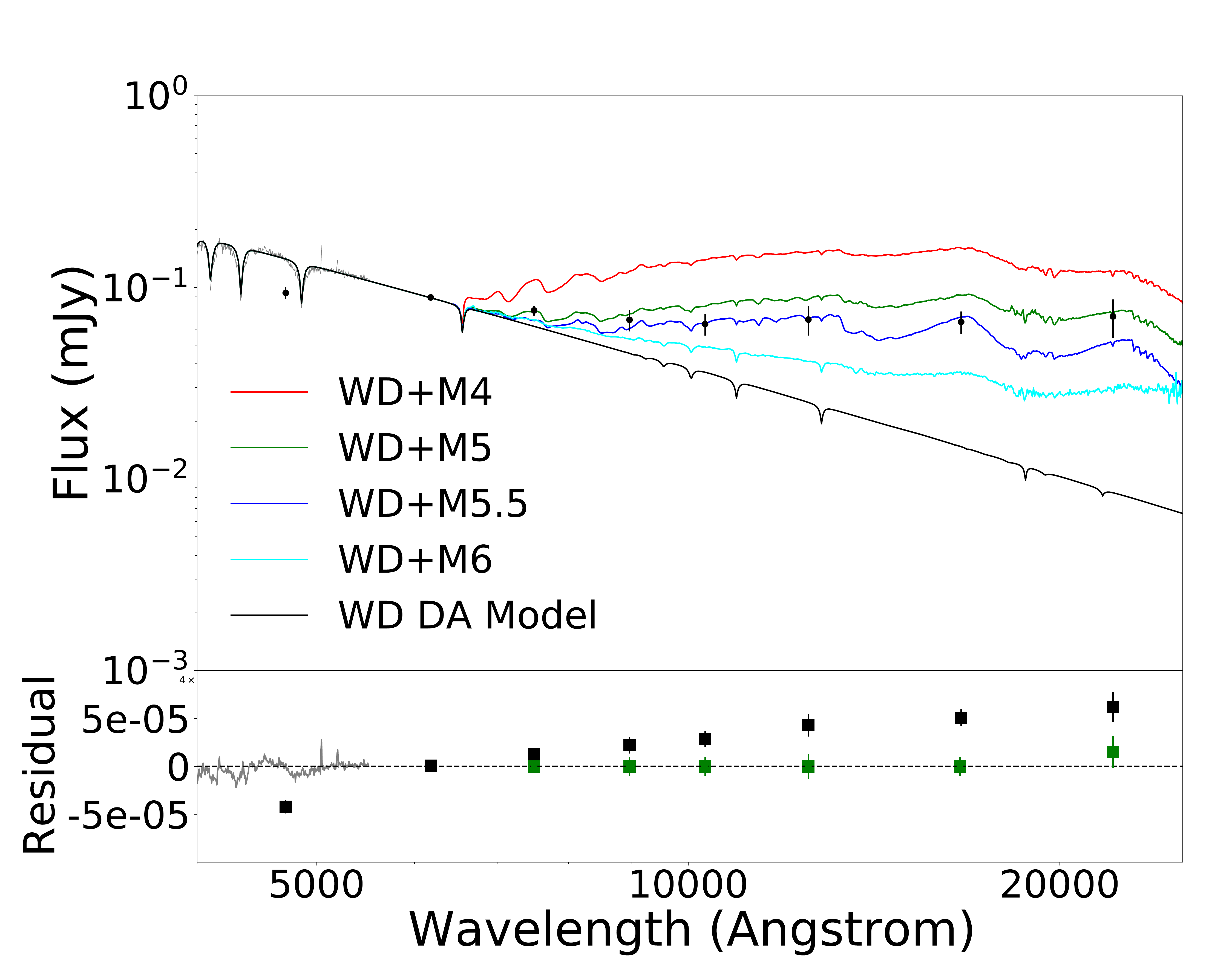}
\caption{OSIRIS spectrum of SDSS103844.58$+$110053.5 shown with the SDSS and UKIDSS photometry and the combined white dwarf model an M5, M5.5 and M6 template spectra. The likely spectral type of the companion is between M5-M6}
\label{fig:1038}
\end{figure}

\begin{figure}
\includegraphics[width=0.9\columnwidth, angle=0]{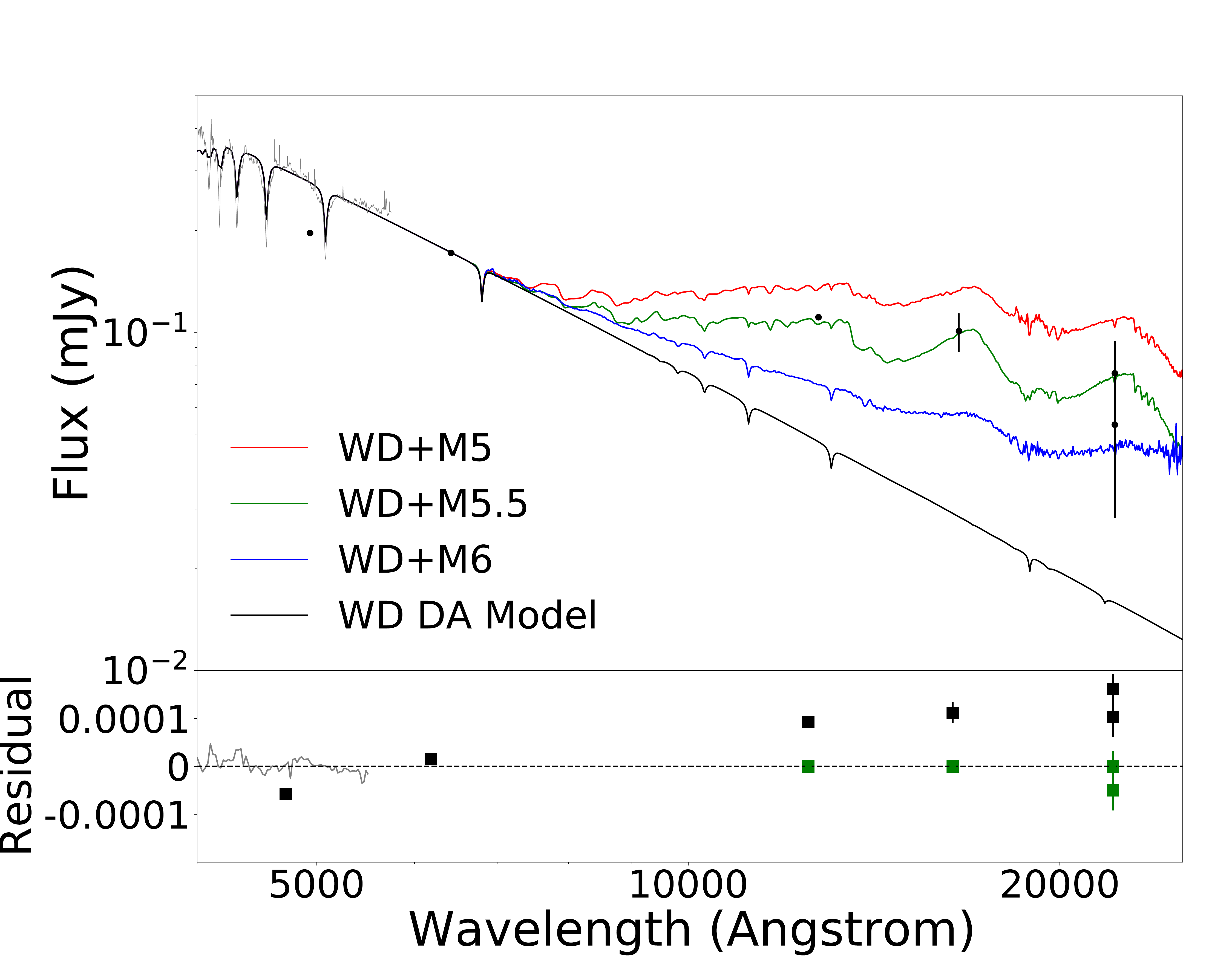}
\caption{OSIRIS spectrum of UVEX184610.80$+$022032.4 shown with the optical and NIR photometry  and the combined white dwarf model and M dwarf template spectra. The suggested spectral type of the companion is M5-M6.}
\label{1846}
\end{figure}

\begin{figure}
\includegraphics[width=0.9\columnwidth, angle=0]{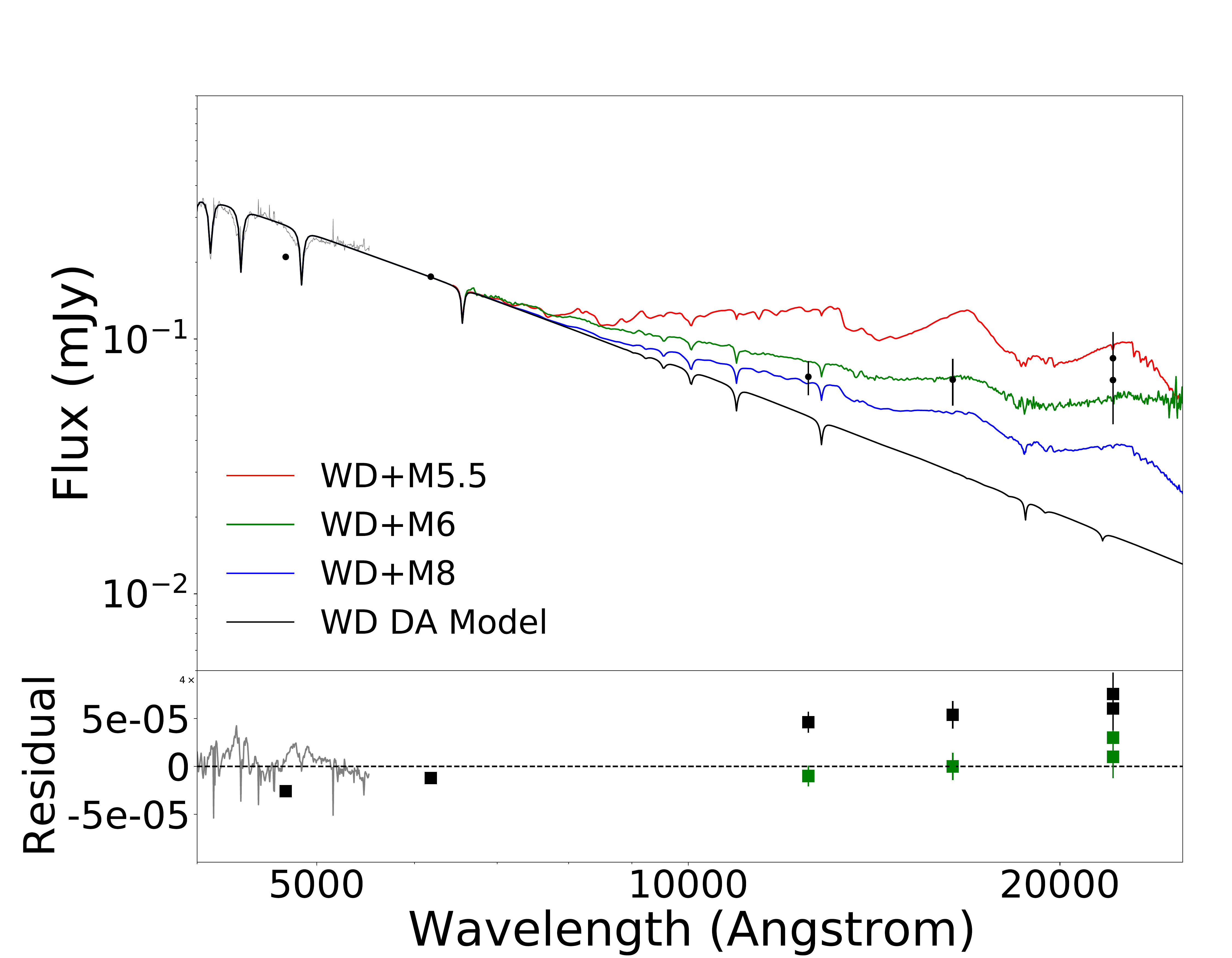}
\caption{OSIRIS spectrum of UVEX185941.43$+$013954.0, shown with SDSS and UKIDSS photometry, the white dwarf model, and the white dwarf+M dwarf combined templates. The photometry indicate the companion is likely to be a M5-M6 dwarf.}
\label{1859}
\end{figure}

\begin{figure}
\includegraphics[width=0.9\columnwidth, angle=0]{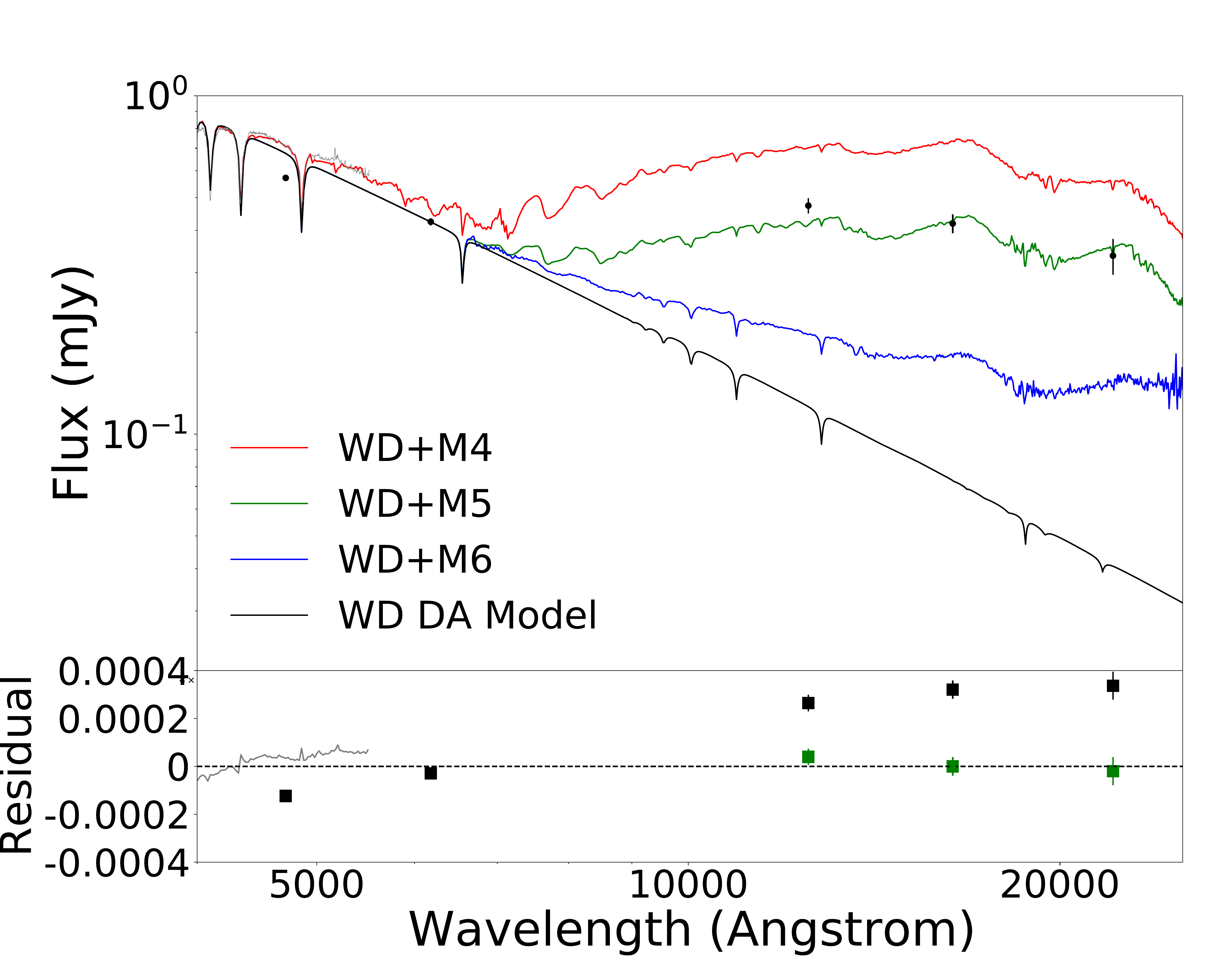}
\caption{OSIRIS spectrum of 2MASS20265915$+$4116436, shown with SDSS and UKIDSS photometry, the white dwarf model, and the white dwarf+brown dwarf combined templates. The photometry indicate the companion is likely to be a M5 dwarf.}
\label{2026}
\end{figure}

\begin{figure}
\includegraphics[width=0.9\columnwidth, angle=0]{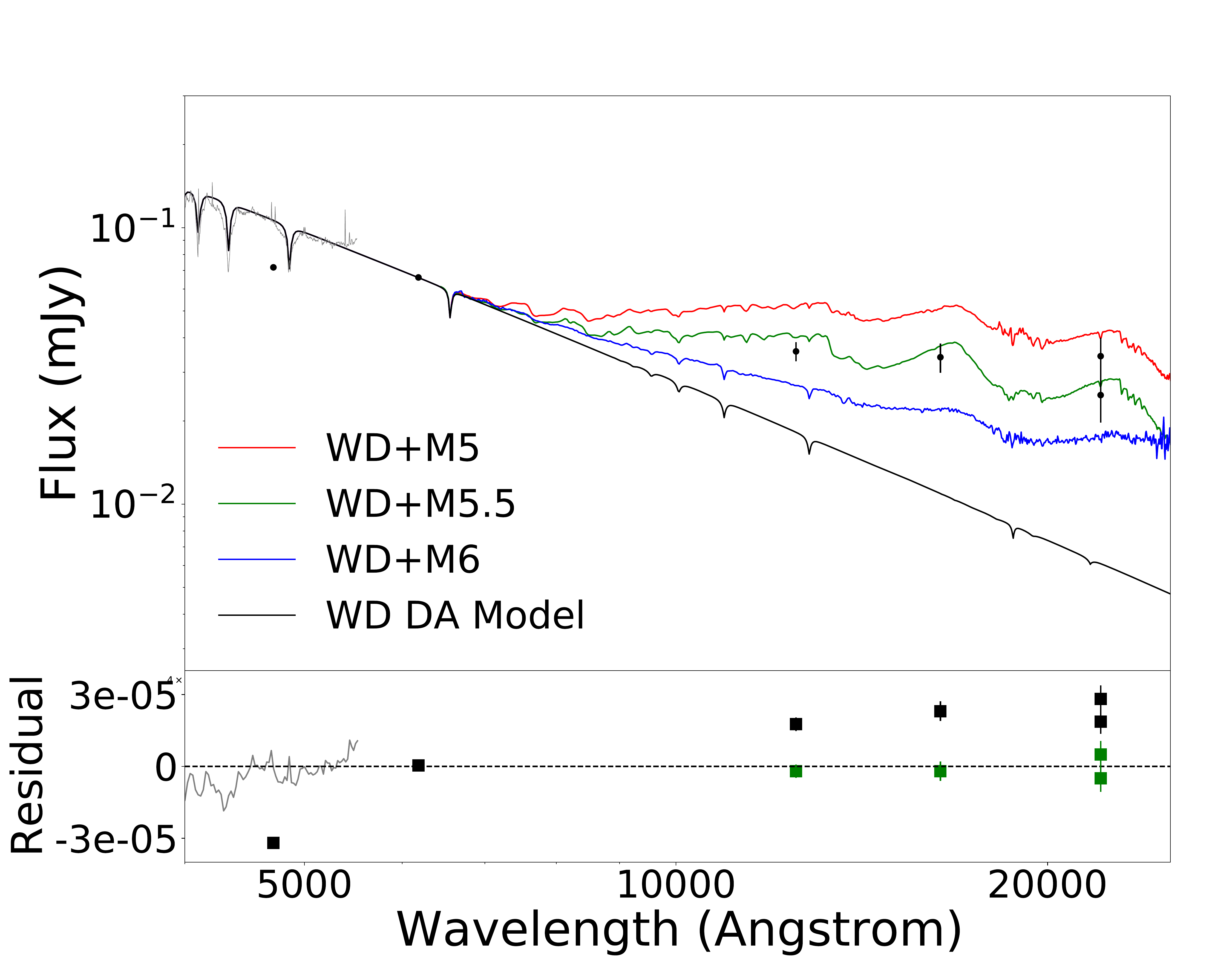}
\caption{OSIRIS spectrum of UVEXJ204229.67$+$384058.0 , shown with SDSS and UKIDSS photometry, the white dwarf model, and the white dwarf+brown dwarf combined templates. The photometry indicate the companion is likely to be a M5-M6 dwarf.}
\label{2042}
\end{figure}

\begin{figure}
\includegraphics[width=0.9\columnwidth, angle=0]{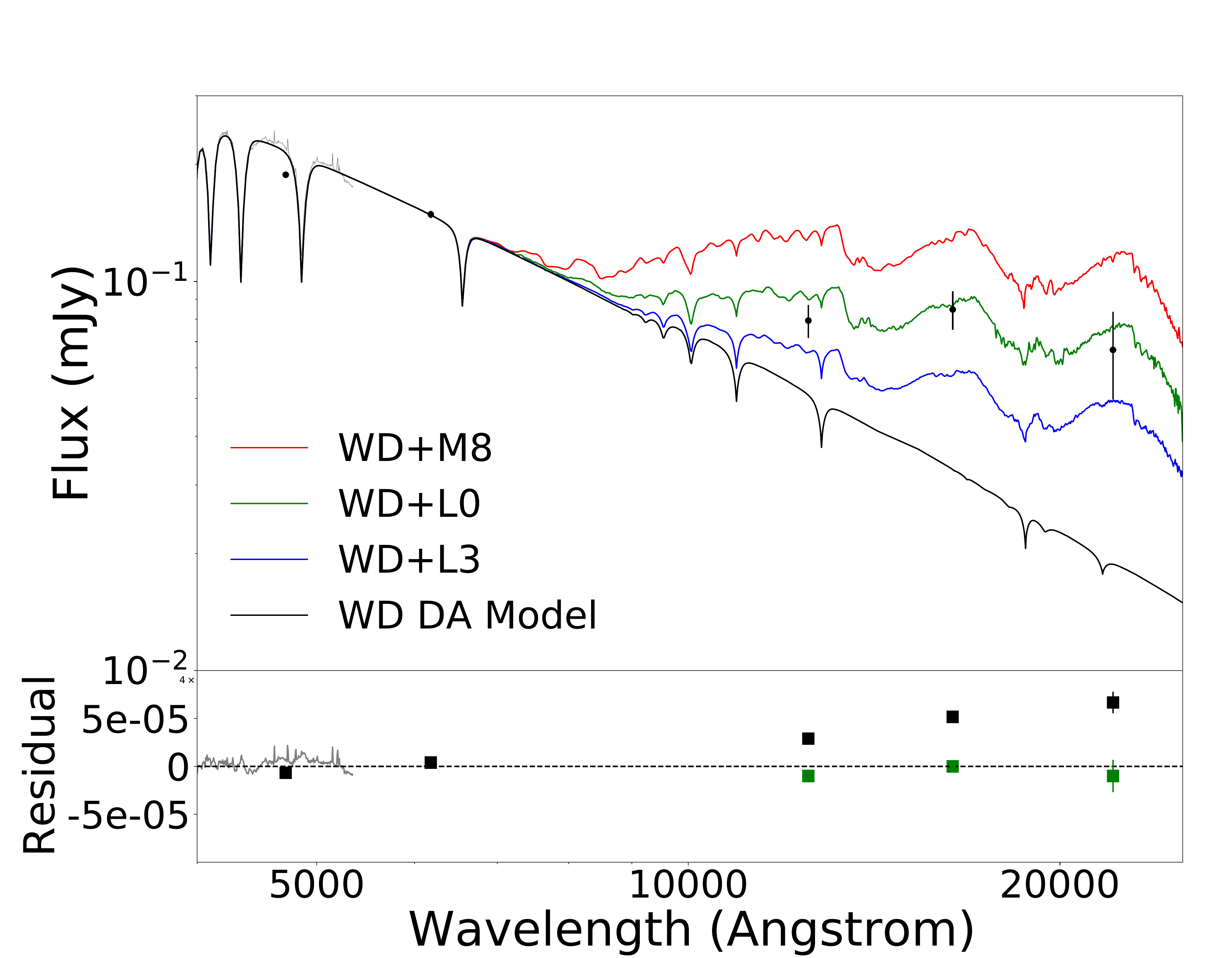}
\caption{OSIRIS spectrum of UGPS210248.46$+$475058.6, shown with SDSS and UKIDSS photometry, the white dwarf model, and the white dwarf+brown dwarf combined templates. The photometry indicate the companion is likely to be M9-L0 dwarf. }
\label{2102}
\end{figure}

%\begin{figure}
%\includegraphics[width=.7\columnwidth, angle=270]{casewell_s_fig4.eps}
%\caption{OSIRIS spectrum of SDSS0751+2002, shown with SDSS and UKIDSS photometry, the white dwarf model, and the white dwarf+brown dwarf combined templates. The photometry indicate the companion is likely to be a L3$-$L6 dwarf.}
%\label{db}
%\end{figure}

% Don't change these lines
\bsp	% typesetting comment
\label{lastpage}
\end{document}